\documentclass[a4paper,fleqn,usenatbib]{mnras}
\usepackage{newtxtext,newtxmath}
\usepackage[T1]{fontenc}
\usepackage{ae,aecompl}
\usepackage{color, soul, ulem,graphicx,commath}
\usepackage{threeparttable}

\def\approxsup{%
  \def\p{%
    \setbox0=\vbox{\hbox{$>$}}%
    \ht0=0.6ex \box0 }%
  \def\s{%
    \vbox{\hbox{$\sim$}}%
  }%
  \mathrel{\raisebox{0.7ex}{%
      \mbox{$\underset{\s}{\p}$}%
    }}%
}

\title[MOVES III]{MOVES III. Simultaneous X-ray and ultraviolet observations unveiling the variable environment of the hot Jupiter HD\,189733b}
\author[V.~Bourrier, et al.]{V.~Bourrier$^{1}$\thanks{E-mail: vincent.bourrier@unige.ch}, P.~J.~Wheatley$^{2,3}$, A.~Lecavelier des Etangs$^4$, G.~King$^{2,3}$, T.~Louden$^{2,3}$,
\newauthor
D.~Ehrenreich$^{1}$, R.~Fares$^5$, Ch.~Helling$^{6,7}$, J.~Llama$^{8}$, M.~M.~Jardine$^{9}$, A.~A.~Vidotto$^{10}$
\\ 
$^{1}$Observatoire de l'Universit\'e de Gen\`eve, Chemin des Maillettes 51, Versoix, CH-1290, Switzerland\\
$^{2}$Department of Physics, University of Warwick, Gibbet Hill Road, Coventry CV4 7AL, UK\\
$^{3}$Centre for Exoplanets and Habitability, University of Warwick, Gibbet Hill Road, Coventry CV4 7AL, UK\\
$^{4}$Institut d’Astrophysique de Paris, CNRS, UMR 7095 \& Sorbonne Universit\'es, UPMC Paris 6, 98 bis bd Arago, 75014 Paris, France\\
$^{5}$Physics Department, United Arab Emirates University, P.O. Box 15551, Al-Ain, United Arab Emirates\\
$^{6}$Centre for Exoplanet Science, School of Physics \&
Astronomy, University of St Andrews, North Haugh, St
Andrews, KY16 9SS\\
$^{7}$SRON Netherlands Institute for Space Research,
Sorbonnelaan 2, 3584 CA Utrecht, NL\\
$^{8}$Lowell Observatory, 1400 West Mars Hill Road, Flagstaff, AZ 86001, USA\\
$^{9}$SUPA, School of Physics and Astronomy, North Haugh, St Andrews, Fife, KY16 9SS, UK\\
$^{10}$School of Physics, Trinity College Dublin, College Green, Dublin-2, Ireland
}

\date{Accepted XXX. Received YYY; in original form ZZZ}

\pubyear{2020}

\hypersetup{draft}

\begin{document}
\label{firstpage}
\pagerange{\pageref{firstpage}--\pageref{lastpage}}
\maketitle

\begin{abstract}

In this third paper of the MOVES (Multiwavelength Observations of an eVaporating Exoplanet and its Star) programme, we combine Hubble Space Telescope far-ultraviolet observations with \textit{XMM-Newton}/\textit{Swift} X-ray observations to measure the emission of HD\,189733 in various FUV lines, and its soft X-ray spectrum. Based on these measurements we characterise the interstellar medium toward HD\,189733 and derive semi-synthetic XUV spectra of the star, which are used to study the evolution of its high-energy emission at five different epochs. Two flares from HD\,189733 are observed, but we propose that the long-term variations in its spectral energy distribution have the most important consequences for the environment of HD\,189733b. Reduced coronal and wind activity could favour the formation of a dense population of Si$^{2+}$ atoms in a bow-shock ahead of the planet, responsible for pre- and in-transit absorption measured in the first two epochs. In-transit absorption signatures are detected in the Lyman-$\alpha$ line in the second, third and fifth epochs, which could arise from the extended planetary thermosphere and a tail of stellar wind protons neutralised via charge-exchange with the planetary exosphere. We propose that increases in the X-ray irradiation of the planet, and decreases in its EUV irradiation causing lower photoionisation rates of neutral hydrogen, favour the detection of these signatures by sustaining larger densities of H$^{0}$ atoms in the upper atmosphere and boosting charge-exchanges with the stellar wind. Deeper and broader absorption signatures in the last epoch suggest that the planet entered a different evaporation regime, providing clues as to the link between stellar activity and the structure of the planetary environment.\\

\end{abstract}

\section{Introduction}\label{sec_intro}

Nearly half of the known exoplanets orbit within 0.1\,au from their star. At such close distances, the nature and evolution of these planets is shaped by interactions with their host star (irradiation, tidal effects, and magnetic fields). In particular, the deposition of stellar X-ray and extreme ultraviolet radiation (XUV) into an exoplanet upper atmosphere can lead to its hydrodynamic expansion and substantial escape (e.g., \citealt{VM2003}, \citealt{Lammer2003}, \citealt{Lecav2004}, \citealt{Yelle2004}, \citealt{GarciaMunoz2007}, \citealt{Koskinen2010}, \citealt{Johnstone2015}). Atmospheric loss is considered one of the main processes behind the deficit of Neptune-mass planets at close orbital distances (the so-called hot Neptune desert, e.g., \citealt{Lecav2007}, \citealt{Davis2009}, \citealt{szabo_kiss2011}, \citealt{Lopez2012,Lopez2013}, \citealt{beauge2013}, \citealt{Owen2013}, \citealt{Kurokawa2014}, \citealt{Jin2014}, \citealt{Lundkvist2016}). These planets are large enough to capture much of the stellar energy, but in contrast to hot Jupiters are not massive enough to retain their escaping atmospheres (e.g., \citealt{Lecav2007}, \citealt{Hubbard2007}, \citealt{Ehrenreich2011}). The missing hot Neptunes could have lost their entire atmosphere via evaporation, evolving into bare rocky cores at the lower-radius side of the desert (e.g. \citealt{Lecav2004}, \citealt{Owen2012}). This scenario is strengthened by the recent observations of warm Neptunes at the border of the desert, on the verge of (\citealt{Kulow2014}, \citealt{Ehrenreich2015}, \citealt{Bourrier2015_GJ436,Bourrier2016}, \citealt{Lavie2017}) or undergoing (\citealt{Bourrier2018_GJ3470b}) considerable mass loss. Because they survive more extreme conditions than lower-mass gaseous exoplanets, hot Jupiters are particularly interesting targets to study star-planet interactions. Their upper atmosphere can be substantially ionised because of stellar photoionisation (e.g. \citealt{Schneiter2016}), which could help the formation of reconnections between the stellar and planetary magnetospheres that would enhance stellar activity (e.g. \citealt{Ip2004}, \citealt{Cuntz2000}, \citealt{Shkolnik2003,Shkolnik2008};  \citealt{Scandariato2013}; although see \citealt{Poppenhaeger2011}; \citealt{Llama2015} for the difficulties to detect such signatures). Atmospheric escape of neutral hydrogen and metal species has been detected via transmission spectroscopy for several Jupiter-mass planets, bringing information about their upper atmosphere and the stellar environment (HD\,209458b, \citealt{VM2003,VM2004,VM2008}, \citealt{Ehrenreich2008}, \citealt{BJ_Hosseini2010}; \citealt{Linsky2010}, \citealt{Schlawin2010}, \citealt{Ballester2015} \citealt{VM2013}; HD\,189733b, \citealt{Lecav2010,Lecav2012}; \citealt{Bourrier2013}; 55\,Cnc\,b, \citealt{Ehrenreich2012}; WASP-12b, \citealt{Fossati2010}, \citealt{Haswell2012}). Shocks could for example form ahead of hot Jupiters because of the interaction between the stellar wind and the planetary outflow or magnetosphere (\citealt{Vidotto2010}, \citealt{Cohen2011}, \citealt{Tremblin2012}, \citealt{Llama2013}, \citealt{Matsakos2015}). \\

The HD\,189733 system offers the possibility to study these various interactions (Table~\ref{tab:param_sys}). It is a binary system with a K2V dwarf (HD\,189733 A, hereafter HD\,189733) and a M4V dwarf (HD\,189733 B, \citealt{Bakos2006}) at a mean separation of $\sim$220\,au. The bright primary (V= 7.7) hosts a transiting hot Jupiter at 0.03\,au (\citealt{Bouchy2005}), whose strong irradiation and large occultation area make it particularly favourable for atmospheric characterisation (see \citealt{Pino2018} and references inside for observations of the lower atmospheric layers). Recent transit observations in the near-infrared (\citealt{Salz2018}) revealed absorption by helium in an extended but compact thermosphere. Transit observations in the far-ultraviolet (FUV) previously revealed absorption by a dense and hot layer of neutral oxygen at higher altitudes (\citealt{BJ_ballester2013}). The close distance of HD\,189733 to the Sun (19.8\,pc) enables observations of the stellar Lyman-$\alpha$ line with the Hubble Space Telescope (\textit{HST}). Atmospheric escape of neutral hydrogen was first detected in the unresolved line with the \textit{HST} Advanced Camera for Surveys (ACS) in two out of three epochs (\citealt{Lecav2010}), and in the line resolved with the \textit{HST} Space Telescope Imaging Spectrograph (STIS) in one out of two epochs (\citealt{Lecav2012}; \citealt{Bourrier2013}). These observations provided the first indication of temporal variations in the physical conditions of an evaporating planetary atmosphere. \citet{Bourrier_lecav2013} attributed the high-velocity, blueshifted absorption signature detected by \citet{Lecav2012} to intense charge-exchange between the planet exosphere and the stellar wind, proposing that an X-ray flare observed before the transit increased the atmospheric mass loss and/or increased the density of the stellar wind. The latter scenario is favoured by thermal escape simulations from \citet{Chadney2017}, who found that the energy input from a flare would not increase sufficiently the mass loss. Interestingly the tentative detection of absorption by ionised carbon (\citealt{BJ_ballester2013}) and excited hydrogen (\citealt{Jensen2012}, \citealt{Cauley2015,Cauley2016,Cauley2017}, \citealt{Kohl2018}) before and during the transit of HD\,189733b could be explained by the interaction of the stellar wind with the planetary magnetosphere or escaping material. The observed temporal variability in the atmospheric escape of neutral and excited hydrogen is likely linked to the high-level of activity from the host star, which results in a fast-changing radiation, particle and magnetic environment for the planet. Enhanced activity in the stellar chromosphere and transition region have also been observed after the planetary eclipse, and attributed to signatures of magnetic star-planet interactions  (\citealt{Pillitteri2010, Pillitteri2011, Pillitteri2014, Pillitteri2015}). Evidence for modulation in the \ion{Ca}{ii} lines at the orbital period of the planet, during an epoch of strong stellar magnetic field, further supports this scenario (\citealt{Cauley2018}). The detectability of star-planet interactions in the HD\,189733 system however remain uncertain, and intrinsic stellar variability and inadequate sampling has been proposed to explain the observed variations (\citealt{Route2019}). These results show the need to study contemporaneously and in different epochs the upper atmosphere of HD\,189733b and its high-energy environment.\\

In this context, we started a multiwavelength observational campaign of this system, in the frame of the MOVES collaboration (Multiwavelength Observations of an eVaporating Exoplanet and its Star, PI V. Bourrier). Observations of the star and the planet were obtained at similar epochs with ground-based and space-borne instruments : in X-rays with the X-ray Multi-Mirror Mission (\textit{XMM-Newton}) and Neil Gehrels Swift Observatory (\textit{Swift}); in the UV with \textit{HST} and \textit{XMM-Newton}; in optical spectropolarimetry with \textit{NARVAL} (\citealt{Auriere2003}) and the Echelle SpectroPolarimetric Device for the Observation of Stars (\textit{ESPaDOnS}, \citealt{Donati2003}, \citealt{Donati2006}); and in radio with the Low-Frequency Array (\textit{LOFAR}, \citealt{vanHaarlem2013}). In MOVES I (\citealt{Fares2017}), we used optical spectropolarimetry of HD\,189733 to reconstruct its surface and large-scale magnetic field in five epochs (2013 Jul, 2013 Aug, 2013 Sept, 2014 Sept and 2015 Jun). We combined these results with data from \citet{Moutou2007}, \citet{Fares2010} to study the evolution of the field over 9 years, and found that its strength changed significantly even though its overall structure remained stable (toroidally-dominated and with the same polarity). We showed that the magnetic environment is not homogeneous over the orbit of HD\,189733b and varies between observing epochs. The resulting inhomogeneities of the stellar wind at the location of the planet (\citealt{Llama2013}) and its variations with the overall stellar magnetic field could cause variations in the structure of the planetary upper atmosphere and its UV transit light curve, on short time-scales (on the order of an orbital period) as well as longer time-scale (on the order of a year). In MOVES II (\citealt{Kavanagh2019}), we thus studied the stellar wind of HD\,189733 and modelled the local particle and magnetic environment surrounding the planet. Our aim was to predict the radio environment of the system (emissions from the stellar wind and planet). From mid-2013 until mid-2015, we showed that, the yearly variation of the stellar wind, together with their inhomogeneities along the planet's orbit, indeed leads to significant variabilities in surrounding medium of HD\,189733b. These results indicate that the best approach to characterise the upper atmosphere of a close-in planet, and to fully understand the physical interactions taking place with its star/stellar wind is to obtain, as much as possible, contemporaneous, multi-wavelength observations of the system. In this paper (MOVES III), we perform a consistent analysis of five transit observations of HD\,189733b in the FUV. We interpret these FUV observations together with X-ray observations of HD\,189733 that were obtained simultaneously for all visits but the first. We present the observations and their analysis in Sect.~\ref{sec:obs_datared}. The search for variations in the Lyman-$\alpha$ line and other FUV stellar lines is described in Sect.~\ref{sec:search_var}. We derive and discuss the temporal evolution of the intrinsic Lyman-$\alpha$ line and XUV spectrum of the star in Sect.~\ref{sec:HE_spec}. Our results are interpreted in terms of stellar evolution, planetary atmospheric escape, and star-planet interactions (SPI) in Sect.~\ref{sec:interp}. Conclusions and perspectives are presented in Sect.~\ref{sec:conc}.

\begin{table*}
\caption{Properties of the HD\,189733 system fixed in our study.}                                                 
\begin{threeparttable}
\begin{tabular}{llccccc}
\hline
\hline
\noalign{\smallskip}
Parameters        & Symbol      & Value          & Reference       \\
\noalign{\smallskip}
\hline
\noalign{\smallskip}
Distance from Earth     & $D_{\mathrm{*}}$    &    19.78$\pm$0.01\,pc & \citealt{Gaia2018}   \\
Star radius             & $R_{\mathrm{*}}$    &    0.780$\stackrel{+0.017}{_{-0.024}}\,R_{\mathrm{\sun}}$ & \citealt{Gaia2018}    \\
Star mass            & $M_{\mathrm{*}}$      &    0.823$\pm$0.029      $\,M_{\mathrm{\sun}}$  & \citealt{Triaud2009}  \\
Heliocentric stellar radial velocity            & $\gamma_{\mathrm{*}/\mathrm{\sun}}$      &   -2.55$\pm$0.16\,km\,s$^{-1}$  & \citealt{Gaia2018}  \\
Planet-to-star radius ratio     & $R_{\mathrm{p}}/R_{\mathrm{*}}$    &  0.1571$\pm$0.0004   & \citealt{Baluev2015}      \\
Orbital period        & $P_{\mathrm{p}}$   &  2.218575200$\pm$7.7$\times$10$^{-8}$\,days   & \citealt{Baluev2015}    \\
Transit centre  & $T_{\mathrm{0}}$                       &   2453955.5255511$\pm$8.8$\times$10$^{-6}$\,BJD$_\mathrm{TDB}$  & \citealt{Baluev2015} \\
Scaled semi-major axis   & $a_{\mathrm{p}}/R_{\mathrm{*}}$  & 8.863$\pm$0.020 & \citealt{Agol2010} \\
Eccentricity       & $e$         &    0  &   \citealt{Bouchy2005}    \\
Argument of periastron  & $\omega$       &    90$^{\circ}$  &    \citealt{Bouchy2005}     \\
Inclination       & $i_{\mathrm{p}}$           &  85.710$\pm$0.024$^{\circ}$   & \citealt{Agol2010} \\
Impact parameter       & $b$     &   0.6636$\pm$0.0019       & \citealt{Baluev2015}  \\
\noalign{\smallskip}
\hline
\hline
\end{tabular}
  \end{threeparttable}
\label{tab:param_sys}
\end{table*}

\section{Observations and data analysis}
\label{sec:obs_datared}

\subsection{HST STIS observations}
\label{sec:obs}

We analysed transit observations of HD\,189733b obtained in five independent epochs with the \textit{HST}/STIS (\citealt{Woodgate1998}). Two archival datasets published in \citet{Lecav2012}, \citet{Bourrier2013} (Visits A and B) are combined with our three original datasets (PI: P.J. Wheatley; Visits C, D, and E). All data were obtained with the \textit{STIS}/G140M grating (spectral range 1195 to 1248\,\AA, spectral resolution $\sim$20\,km\,s$^{-1}$), with the main purpose of searching for the transit of HD\,189733b in the stellar Lyman-$\alpha$ line. The log of the observations is given in Table~\ref{tab:log}. Visits A, B, and C each consist of four consecutive \textit{HST} orbits obtained before, during, and after the planetary transit. Visits D and E each consist of three consecutive \textit{HST} orbits obtained before and during the transit. Data obtained in time-tagged mode were reduced with the \textsc{calstis} pipeline (version 3.4, \citealt{Hodge1998}), which includes the flux and wavelength calibration, and divided in sub-exposures in each \textit{HST} orbit. The first orbit (orbit 1) in each visit has a shorter scientific exposure because of target acquisition. In Visits A and B, acquisition-peakup exposures were performed at the start of orbits 2-3-4 in case the target needed recentring within the slit. This operation was no longer performed in Visits C to E, which further had the wavelength calibration exposures carried out during Earth occultation. This resulted in substantially longer scientific exposures in orbits 2-3-4 for those visits. The number of sub-exposures in the orbits of each visit was thus adjusted to keep a duration of $\sim$300\,s\\

\begin{table*}
\centering
\begin{minipage}[tbh]{\textwidth}
\caption{Log of HD\,189733 \textit{HST}, \textit{XMM-Newton}, and \textit{Swift} observations}
\centering
\begin{threeparttable}
\begin{tabular}{lcccccccc}
\hline
\hline
\noalign{\smallskip}	
					&	Date &	Telescope   & \multicolumn{2}{c}{Time from mid-transit (UT)}	 & \multicolumn{2}{c}{Time (BJD$_\mathrm{TDB}$ - 2450000 )}	  &	 Number of  & Duration of   \\	
					&    &    &  Start      &    End   &  Start      &    End          &	sub-exposures & 	sub-exposures (s)	  \\	
\noalign{\smallskip}
\hline
\noalign{\smallskip}
Visit A & 2010-04-06   &  \textit{HST}   & -03:17:45 & 02:00:53    & 5293.18907   & 5293.41035   & 6-7-7-7	 & 316-322-322-322	 \\
\hline
Visit B & 2011-09-07/08 &  \textit{HST}  & -03:25:38 & 01:52:48  & 5812.33019   & 5812.55133  & 6-7-7-7	 & 316-322-322-322	 \\
        & 			   &  \textit{Swift}  & -15:19:48 & 11:53:54  & 5811.83424  & 5812.96875     & 	16 &  572-1667	 \\
\hline
Visit C & 2013-05-09/10 &  \textit{HST}  & -02:30:37 & 02:39:28  & 6422.47658   & 6422.69191  & 5-10-10-10	 & 277-293-293-293	 \\
        & 			   &  \textit{XMM-Newton}  & -05:32:22  & 05:06:48  & 6422.35036  & 6422.79422  & 41	 & 	970 \\
\hline
Visit D & 2013-11-03 &  \textit{HST}  & -02:48:39 & 00:51:52  & 6599.95008   &  6600.10320 & 6-10-10	  & 314-299-299	\\
        & 			   &  \textit{XMM-Newton}  & -05:17:06 & 04:30:54  &  6599.84698 & 6600.25531  & 36	 & 	970 \\
\hline
Visit E & 2013-11-21 & \textit{HST}   & -02:59:01 & 00:41:42  &  6617.69148  & 6617.84475  & 6-10-10	  & 314-299-299	\\
        & 			   &  \textit{XMM-Newton}  & -06:09:10 &  05:00:10  & 6617.55943  &  6618.02424 & 	43 & 	970 \\
\noalign{\smallskip}
\hline
\hline
\end{tabular}
\begin{tablenotes}[para,flushleft]
Notes: The number and duration of sub-exposures are given for each \textit{HST} orbit in \textit{STIS} data. For \textit{Swift}, sub-exposure correspond to individual observations taken on different spacecraft orbits. For \textit{XMM-Newton}, individual exposures are very short and the values indicated correspond to our choice of binning. No X-ray data was obtained in Visit A. 
  \end{tablenotes}
  \end{threeparttable}
\label{tab:log}
\end{minipage}
\end{table*}

\begin{figure*}     
\begin{minipage}[tbh]{\textwidth}
\includegraphics[trim=0cm 0cm 0cm 0cm,clip=true,width=\columnwidth]{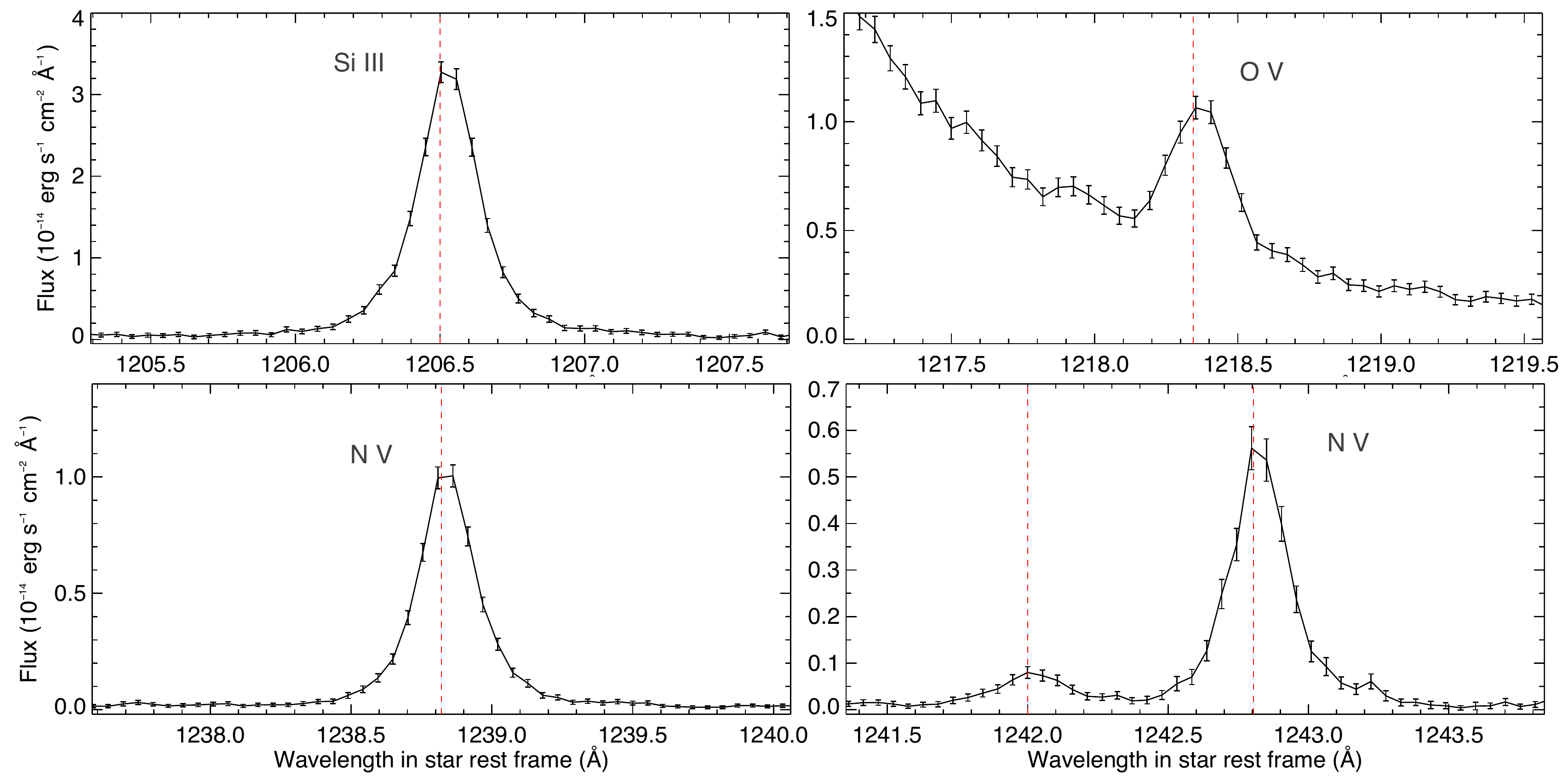}
\caption[]{Spectral profiles of HD\,189733 secondary FUV lines, averaged over the quiescent spectra in all visits. Vertical dashed red lines show the wavelengths of the transitions associated to each stellar line in the star rest frame.}
\label{fig:Fig_stlines}
\end{minipage}
\end{figure*}

\begin{figure}     
\includegraphics[trim=0cm 0cm 0cm 0cm,clip=true,width=0.95\columnwidth]{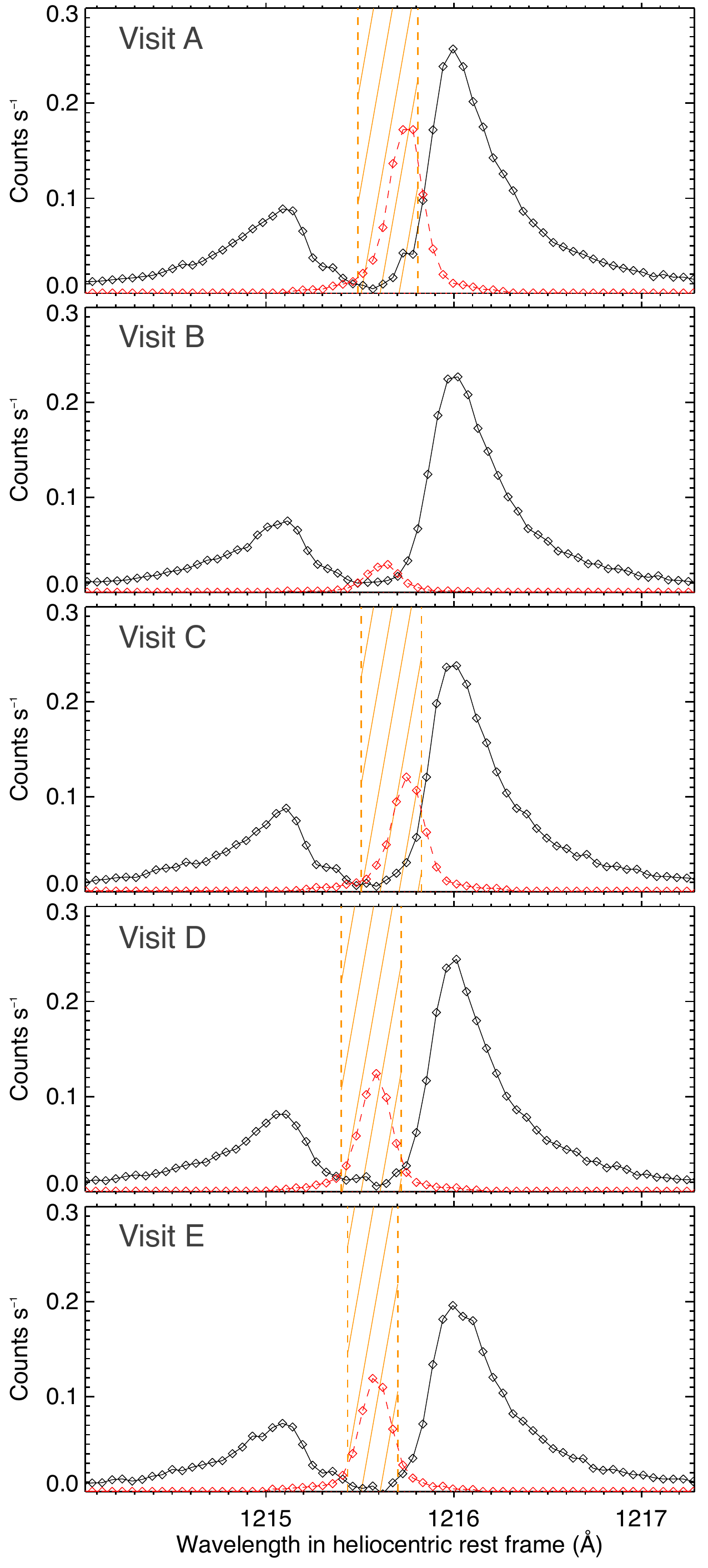}
\caption[]{Raw spectra of HD\,189733 Lyman-$\alpha$ line (black profiles) after correction for the geocoronal emission line (red profiles). Spectra are averaged over all orbits in each visit. Dashed orange regions show wavelengths window excluded from our analyses because of strong airglow contamination.}
\label{fig:Fig_airglow}
\end{figure}

A K-type star like HD\,189733 has no measurable continuum emission in the G140M spectral range. As can be seen in Fig.~\ref{fig:Fig_stlines}, we identified in each visit the following stellar emission lines: Lyman-$\alpha$ line (1215.67\,\AA), \ion{Si}{iii} (1206.5\,\AA), \ion{O}{v} (1218.3\,\AA), \ion{Fe}{xii} (1242.0\,\AA), and the \ion{N}{v} doublet (1242.8\,\AA\, and 1238.8\,\AA). The stellar Lyman-$\alpha$ line in the raw data is contaminated by geocoronal airglow emission from the upper atmosphere of Earth (\citealt{VM2003}). \textsc{calstis} corrects the final 1D spectra for airglow contamination, but it is recommended to treat with caution the regions where the airglow is stronger than the stellar flux (e.g. \citealt{Bourrier2017_HD976,Bourrier2018_GJ3470b}). The strength and position of the airglow varies with the epoch of observation, and after preliminary analyses of the Lyman-$\alpha$ spectra we identified the wavelength windows shown in Fig.~\ref{fig:Fig_airglow} as unreliable. Airglow is low enough in Visit B that the full stellar line profile could be analysed.

\subsection{HST STIS calibrations}
\label{sec:cal}

\begin{figure}     
\includegraphics[trim=0cm 0cm 0cm 0cm,clip=true,width=\columnwidth]{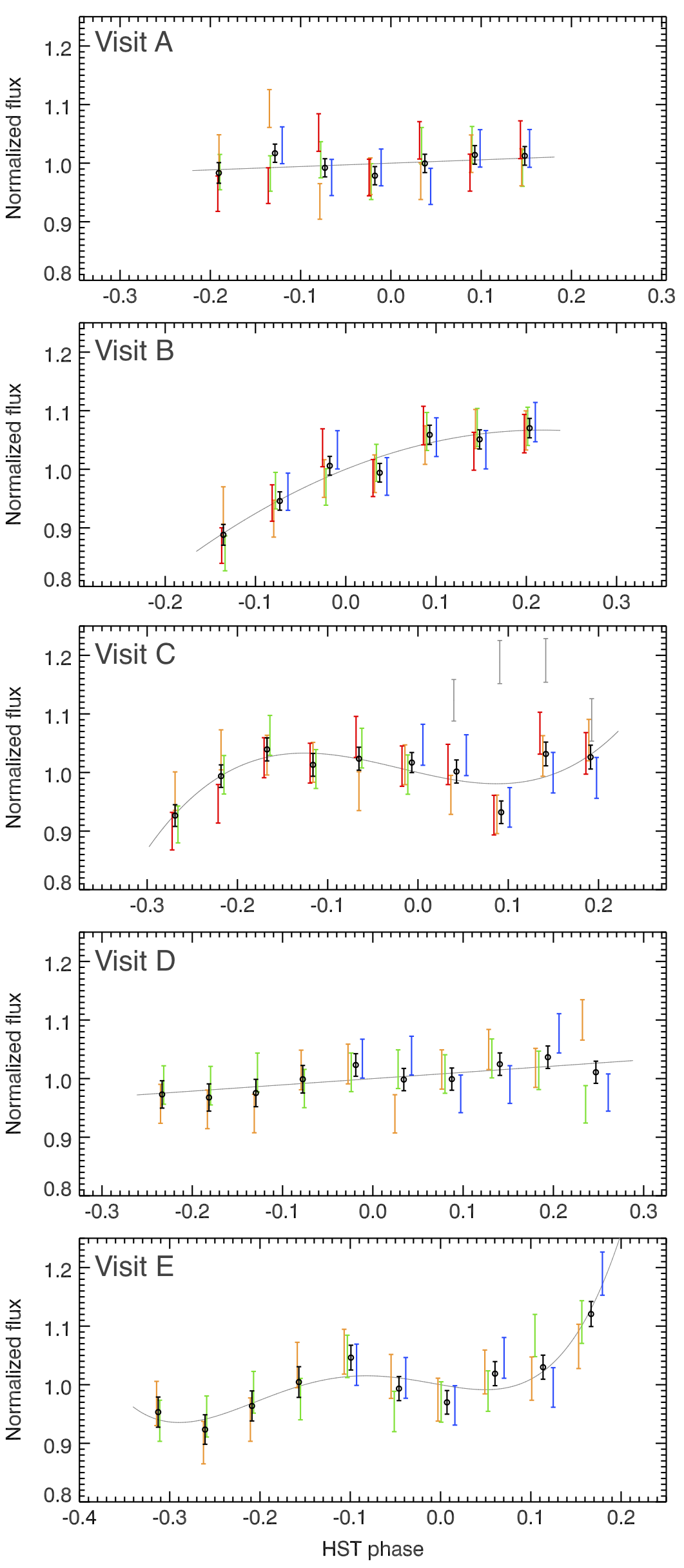}
\caption[]{Lyman-$\alpha$ fluxes for sub-exposures integrated over the entire line and phase-folded on the \textit{HST} orbital period ($P_\mathrm{HST}$ = 96\,min). Phase is between -0.5 and 0.5. Solid grey lines correspond to the best-fit breathing model to the data, which have been scaled to the same nominal flux for the sake of comparison. Black points show sub-exposures binned manually to highlight the breathing trend. Colours correspond to \textit{HST} consecutive orbits in each visit (orbits 1 to 4 are plotted in blue, green, orange, red). The four grey sub-exposures in Visit C are affected by a flare and have been excluded from the fit.}
\label{fig:Fig_breathing}
\end{figure}

The \textit{HST} experiences thermal variations over its orbit, which induce variations in the telescope throughput and modify the balance of the flux measured with \textit{STIS} in each orbit (e.g. \citealt{Brown2001}; \citealt{Sing2008a}; \citealt{Huitson2012}). This ``breathing'' effect is detected in all visits (Fig.~\ref{fig:Fig_breathing}). As in previous measurements with the G140M grating (e.g. \citealt{Bourrier2013}; \citealt{Ehrenreich2015}; \citealt{Bourrier2017_HD976}), the shape and amplitude of the breathing variations change between visits but the orbit-to-orbit variations within a single visit are both stable and highly repeatable, allowing for an efficient correction. Particular care must be taken, however, with the first orbit. Various operations (Sect.~\ref{sec:obs}) make its scientific exposure shorter and shifted to later \textit{HST} orbital phases compared to subsequent exposures (Fig.~\ref{fig:Fig_breathing}). As a result, the flux unbalanced by breathing variations has a different average over the first orbit compared to later orbits. By accounting for this bias we improve on the correction performed for Visits A and B by \citet{Lecav2012} and \citet{Bourrier2013}, who assumed the breathing did not change the average flux over each orbit. \\
We fitted a breathing model based on \citet{Bourrier2017_HD976} to the sub-exposure spectra integrated over the entire Lyman-$\alpha$ line (1214.0--1217.3\,\AA\, minus the range contaminated by the airglow). This choice is motivated by the achromaticity of the breathing variations, and by the need for a high signal-to-noise ratio (SNR) to ensure an accurate correction. The breathing was modelled as a polynomial function phased with the period of the \textit{HST} around the Earth ($P_{\mathrm{HST}}$ = 96\,min). The nominal flux unaffected by the breathing effect was allowed to vary for each \textit{HST} orbit, to prevent the overcorrection of putative orbit-to-orbit variations caused by the star or the planet. We nonetheless excluded from the fit sharp flux variations caused by a flare at the end of orbit 2 in Visit C (Sect.~\ref{sec:VC}). The breathing model was oversampled in time and averaged within the time window of each sub-exposure before comparison with the data. We used the Bayesian Information Criterion (BIC, \citealt{Liddle2007}) as a merit function to determine the best polynomial degree for the breathing variations. The best-fit models, shown in Fig.~\ref{fig:Fig_breathing}, were obtained for degrees of 1, 2, 3, 1, and 4 in visits A to E, respectively. Spectra in each sub-exposure were corrected by the value of the best-fit breathing function at the time of mid-exposure. \\
After correcting the wavelength tables of the spectra for the heliocentric radial velocity of HD\,189733 (Table~\ref{tab:param_sys}), we found that some of the stellar lines were redshifted with respect to their expected rest wavelength relative to the star, as previously noted by \citet{Bourrier2013}. Contrary to these authors we concluded that this redshift has a stellar rather than instrumental origin (see Sect.~\ref{sec:HE_spec}), which is why the velocity ranges we report hereafter (defined in the star rest frame without further correction) are slightly different than in \citet{Lecav2012} and \citet{Bourrier2013}.

\subsection{XMM-Newton and Swift observations}
\label{sec:xray_obs}

We analysed spectra and light curves from three observations of HD\,189733 taken with the \textit{EPIC-pn} camera (\citealt{Struder2001}) onboard \textit{XMM-Newton} in 2013 (ObsID: 0692290201, 0692290301, 0692290401; PI: Wheatley), contemporaneous with \textit{HST} visits C, D and E. We also looked at data from the Optical Monitor, but for the sole purpose of flare identification, and a full analysis will be presented in an independent paper. The log of the observations is given in Table~\ref{tab:log}. The observations were made with the thin optical blocking filter in order to maximise the response to soft X-rays, and in small window mode in order to avoid pile-up. The source was very strongly detected in all three observations. The data were reduced in the standard way using the Scientific Analysis System (\textsc{sas} 16.0.0).

\begin{figure*}     
\includegraphics[trim=0cm 0cm 0cm 0cm,clip=true,width=2.1\columnwidth]{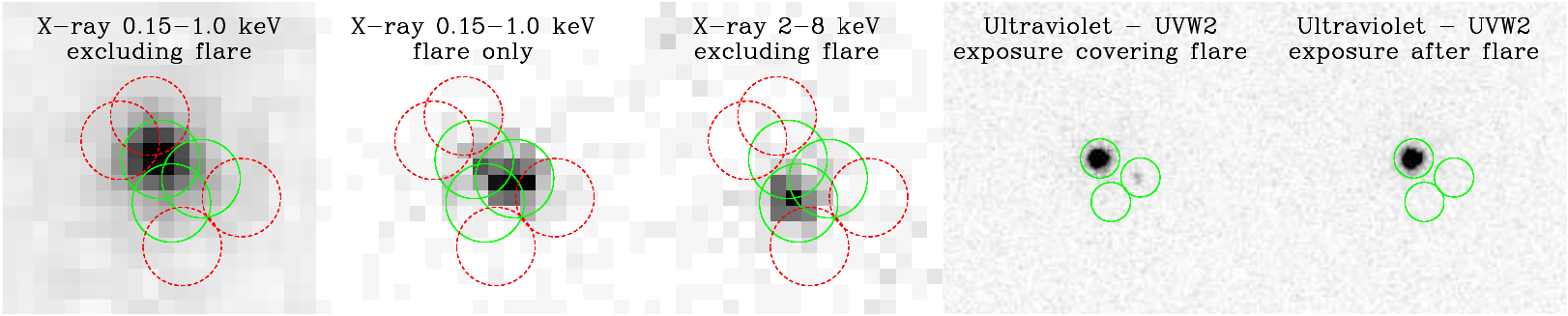}
\caption[]{Plot showing the extraction regions for assessing contributions from the three X-ray sources (HD\,189733, top left; HD\,189733B, right; background X-ray source, bottom left) overplotted on various images from the \textit{XMM-Newton} visit D observations. Green circles show the source extraction regions, and red circles the regions used to estimate the contamination of each source by the two others (see text). The first three plots show \textit{EPIC-pn} data, the last two plots show \textit{OM} data.}
\label{fig:regions}
\end{figure*}

There are two other X-ray sources near to HD\,189733 on the sky, as identified with \textit{Chandra} \citep{Poppenhaeger2013}: the companion M dwarf HD\,189733B, and a background source. The three form a roughly equilateral triangle on the sky, with angular separations of about 12\,arcsec. Fig.~\ref{fig:regions} shows how the contamination of each of the sources by the others was considered. Using small source extraction regions of 10\,arcsec radius (green circles in Fig.~\ref{fig:regions}), we subtracted equivalently-sized regions from the opposite side of the contaminating sources (red dashed circles to estimate the count rate of each component separately. This analysis showed that the contribution of HD\,189733B is negligible at all energies except during a single flaring period in Visit D, which we excluded from our analyses (see Sect.~\ref{sec:search_var}). HD\,189733 and the background source are spatially resolved in \textit{Chandra} observations published by \citet{Poppenhaeger2013}. The comparison of their spectra show that the contribution of the background source is negligible at energies below about 1.2\,keV, where HD\,189733 emits most of its X-ray energy (Sect.~\ref{sec:X_spec}). Therefore we used the larger 15\,arcsec regions to extract the total count rate, and excluded energies above 1.2\,keV to characterise the X-ray emission of HD\,189733 (the \textit{EPIC-pn} camera observes from 0.16 to 15\,keV). We note that most of the X-ray flux is emitted at the softer energies within the 0.166-1.2\,keV energy range (see spectrum in Fig.~\ref{fig:Xspec}), and no significant difference are observed in the variations of the integrated X-ray flux over time when including or excluding harder energies. \\   

We also analysed a set of observations of HD\,189733 taken with the \textit{XRT} instrument (\citealt{Burrows2005}) on \textit{Swift} (ObsID: 00036406010 to 00036406017; PI: Wheatley), simultaneous with \textit{HST} visit B. These observations were previously presented by \citet{Lecav2012}, who identified an X-ray flare about 8 hours before the primary transit of the planet. The \textit{XRT} instrument observes from 0.2 to 10\,keV. As with the \textit{XMM-Newton} data the background source could contaminate the flux at high energies, which were excluded from our analysis of HD189733. For these data we used source and background regions of radius 30 and 100\,arcsec, respectively. These were extracted using the \textsc{xselect} program \footnote{https://heasarc.gsfc.nasa.gov/ftools/xselect/}

\section{Search for FUV spectral line variations}
\label{sec:search_var}

We searched for flux variations in the lines of HD\,189733, which would arise from absorption by the planetary atmosphere or from stellar activity. Spectra were first compared two by two in each visit to identify those showing no significant variations, which could be considered as representative of the quiescent, unocculted stellar lines. This was done by searching for all features characterised by flux variations with S/N larger than 3, and extending over more than 3 pixels ($\sim$0.16\,\AA\,, larger than \textit{STIS}/G140M spectral resolution). For the brightest lines (Lyman-$\alpha$, \ion{Si}{iii}, and the co-added lines of the \ion{N}{v} doublet), we compared spectra averaged not only over each orbit but also over groups of sub-exposures. Once stable spectra were identified for each line, they were coadded into a master quiescent spectrum for each visit, which was used to characterise the features detected in the variable spectra more precisely. We present the results of these analyses in the following sections, along with the X-ray light curves measured in Visits B-E to help disentangling the stellar and planetary variations. The \textit{Swift} light curve (visit B) covers the energy range 0.3 to 1.2\,keV, and the \textit{XMM-Newton} light curves (visits C to E) the energy range 0.16 to 1.2\,keV. \\

The \ion{O}{v} line was analysed after correcting the spectra for the red wing of HD\,189733 Lyman-$\alpha$ line using a polynomial model specific to each visit (Fig.~\ref{fig:Fig_stlines}). However the corrected \ion{O}{v} line is too faint to be analysed spectrally, and we found no significant variations in the flux integrated over the entire line in any of the visits. Therefore we do not discuss variations in the \ion{O}{v} hereafter.\\


\subsection{Visit A - April 6, 2010}
\label{sec:VA}

\begin{figure}     
\includegraphics[trim=0cm 0cm 0cm 0cm,clip=true,width=\columnwidth]{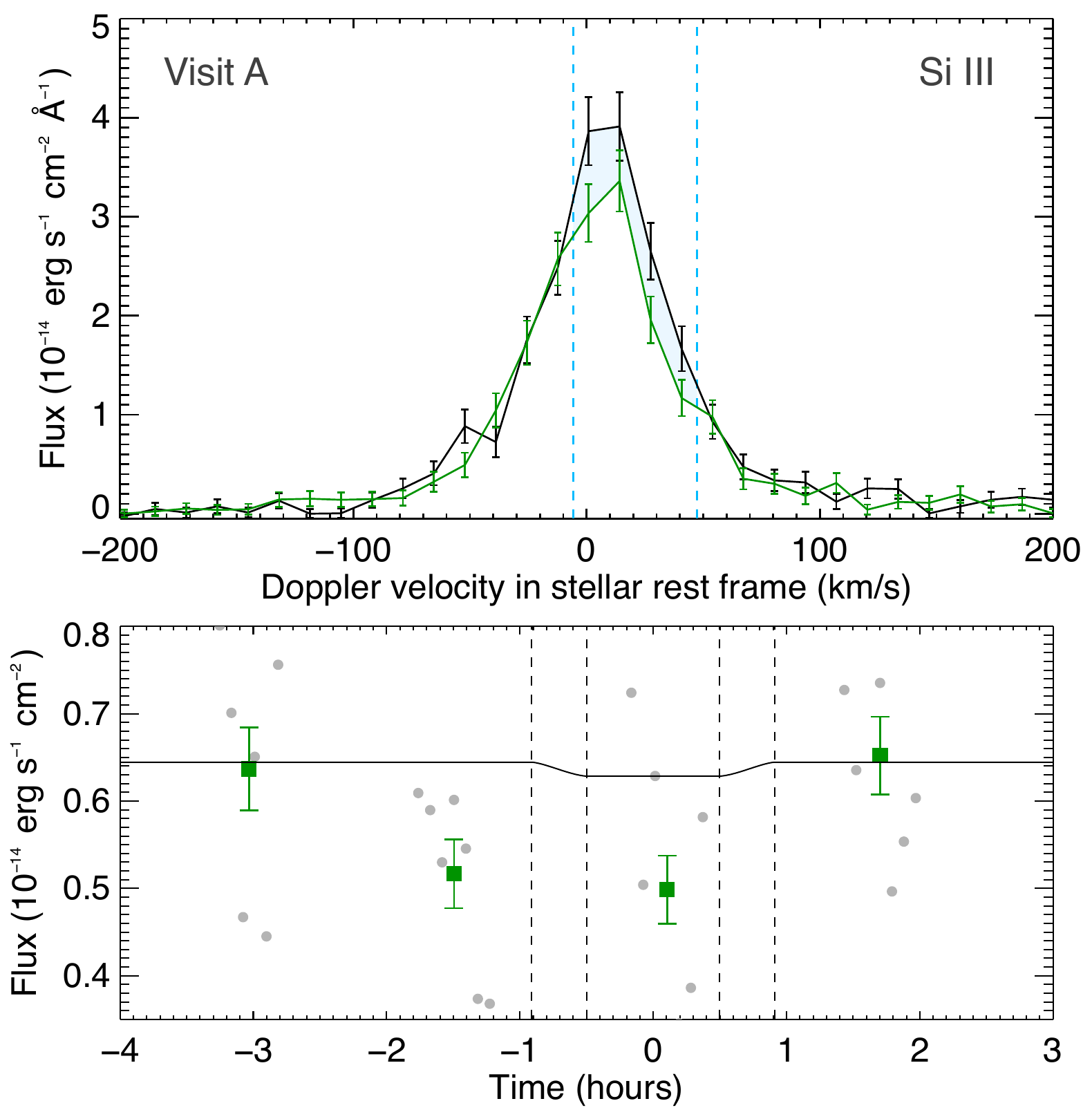}
\caption[]{\ion{Si}{iii} line variations in Visit A. \textit{Top panel: } Stellar line profile, averaged over orbits 1 and 4 (black) and over orbits 2 and 3 (green). The spectral range showing absorption is highlighted in blue. \textit{Bottom panel: } Flux integrated over the absorbed spectral range, as a function of time relative to the planet transit. Green squares correspond to entire orbits, while grey disks stand for sub-exposures. The solid black line is the optical planetary light curve, with contacts shown as dashed vertical lines.}
\label{fig:Fig_SiIII_paper_VA}
\end{figure}

The \ion{Si}{iii} line shows a similar profile in orbits 1 and 4, which is different from the similar profile it shows in orbits 2 and 3 (Fig.~\ref{fig:Fig_SiIII_paper_VA}). We consider that the first group is representative of the intrinsic stellar line because its out-of-transit exposures show a symmetrical line profile. In contrast, mirroring the line in orbits 2 and 3 reveals that it is distorted and misses flux in its peak and red wing. Compared to orbits 1+4 this corresponds to an absorption of 21.2$\pm$5.8\% within -5.7 to 47.4\,km\,s$^{-1}$, which occurs in exposures obtained just before and during the planetary transit (Fig.~\ref{fig:Fig_SiIII_paper_VA}). This localised absorption signature was not detected by \citet{Bourrier2013}, who grouped pre-transit observations and focused on variations over the entire \ion{Si}{iii} line. Other parts of the \ion{Si}{iii} line remain stable during the visit. 

We do not detect any significant variations in the \ion{N}{v} lines, although we note that their wings are marginally brighter in the first orbit compared to subsequent exposures. 

\begin{figure}    
\includegraphics[trim=0cm 0cm 0cm 0cm,clip=true,width=\columnwidth]{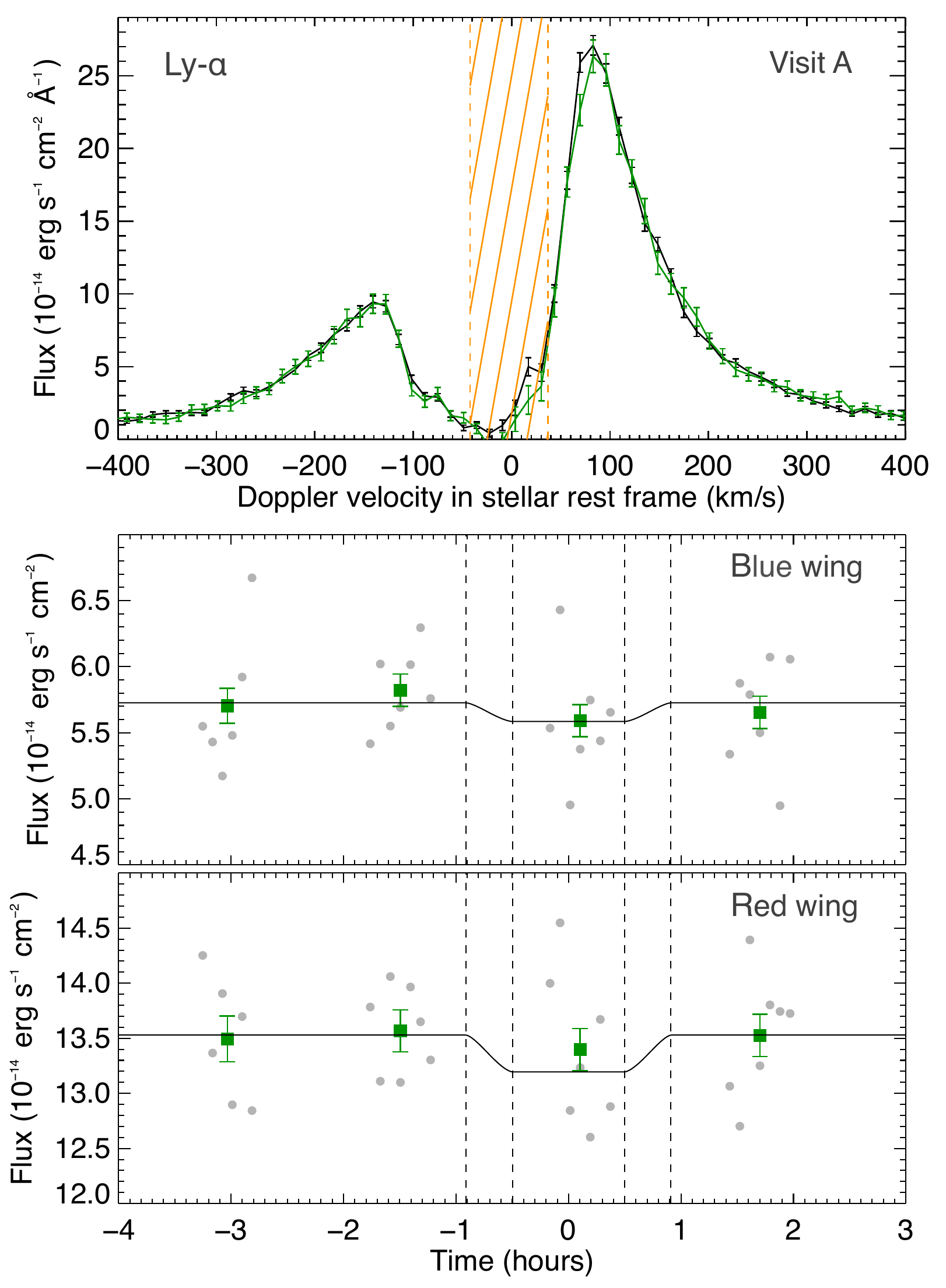}
\caption[]{\textbf{Top panel:} Stellar Lyman-$\alpha$ line in Visit A, during the optical transit of HD\,189733b (green spectrum) and averaged over pre- and post-transit exposures (black spectrum). No variations are detected. The dashed orange region is too contaminated by geocoronal emission to be studied. \textbf{Bottom panels:} Lyman-$\alpha$ flux integrated over the entire blue and red wings of the line, with the same codes as in Fig.~\ref{fig:Fig_SiIII_paper_VA}.}
\label{fig:Fig_Ly_paper_VA}
\end{figure}

We also do not detect any significant variations in the Lyman-$\alpha$ line, in agreement with \citet{Lecav2012}, \citet{Bourrier2013}, and \citet{Guo2016}. We show in Fig.~\ref{fig:Fig_Ly_paper_VA} the comparison between the master out-of-transit spectrum and that obtained during the optical transit, when absorption by an extended exosphere of neutral hydrogen is expected to be strongest (\citealt{Bourrier_lecav2013}).\\

\subsection{Visit B - September 7/8, 2011}
\label{sec:VB}

\begin{figure}     
\includegraphics[trim=0cm 0cm 0cm 0cm,clip=true,width=\columnwidth]{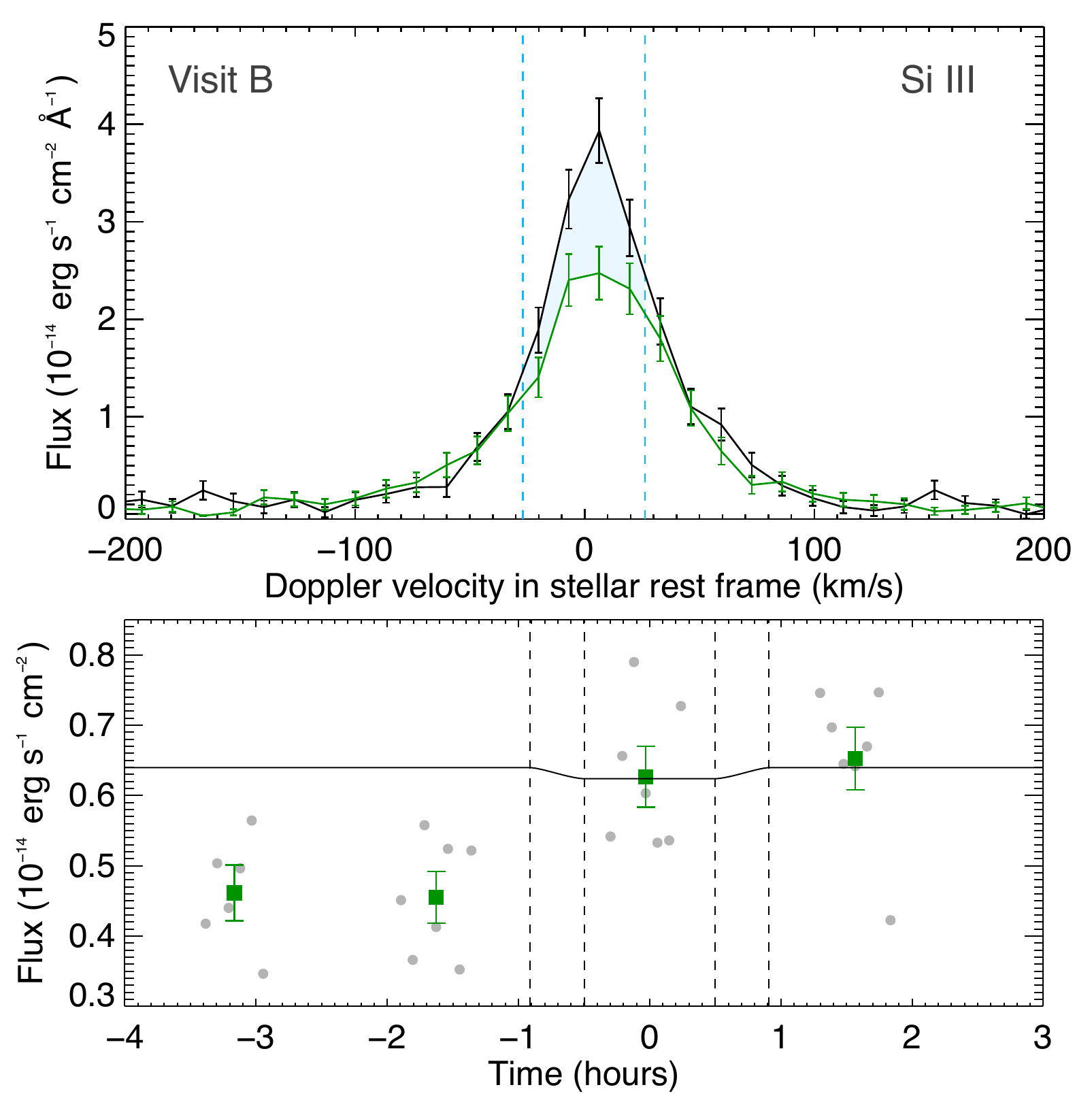}
\caption[]{\ion{Si}{iii} line variations in Visit B. Same description as in Fig.~\ref{fig:Fig_SiIII_paper_VA}, except that the stellar line profile is averaged over orbits 3 and 4 (black) and over orbits 1 and 2 (green).}
\label{fig:Fig_SiIII_paper_VB}
\end{figure}

The \ion{Si}{iii} line shows a similar profile in orbits 1 and 2, which is different from the similar profile it shows in orbits 3 and 4 (Fig.~\ref{fig:Fig_SiIII_paper_VB}). We consider the second group to be representative of the stellar \ion{Si}{iii} line, because its profile is nearly identical to that of the quiescent line in Visit A. In contrast the core of the line shows significant absorption in orbits 1+2 (28.3$\pm$5.5\% within -26.8 to 26.3\,km\,s$^{-1}$), which might extend further in the red wing. These variations were reported by \citet{Bourrier2013}, and occur in the two orbits before the planetary transit (Fig.~\ref{fig:Fig_SiIII_paper_VB}).

\begin{figure}     
\includegraphics[trim=0cm 0cm 0cm 0cm,clip=true,width=\columnwidth]{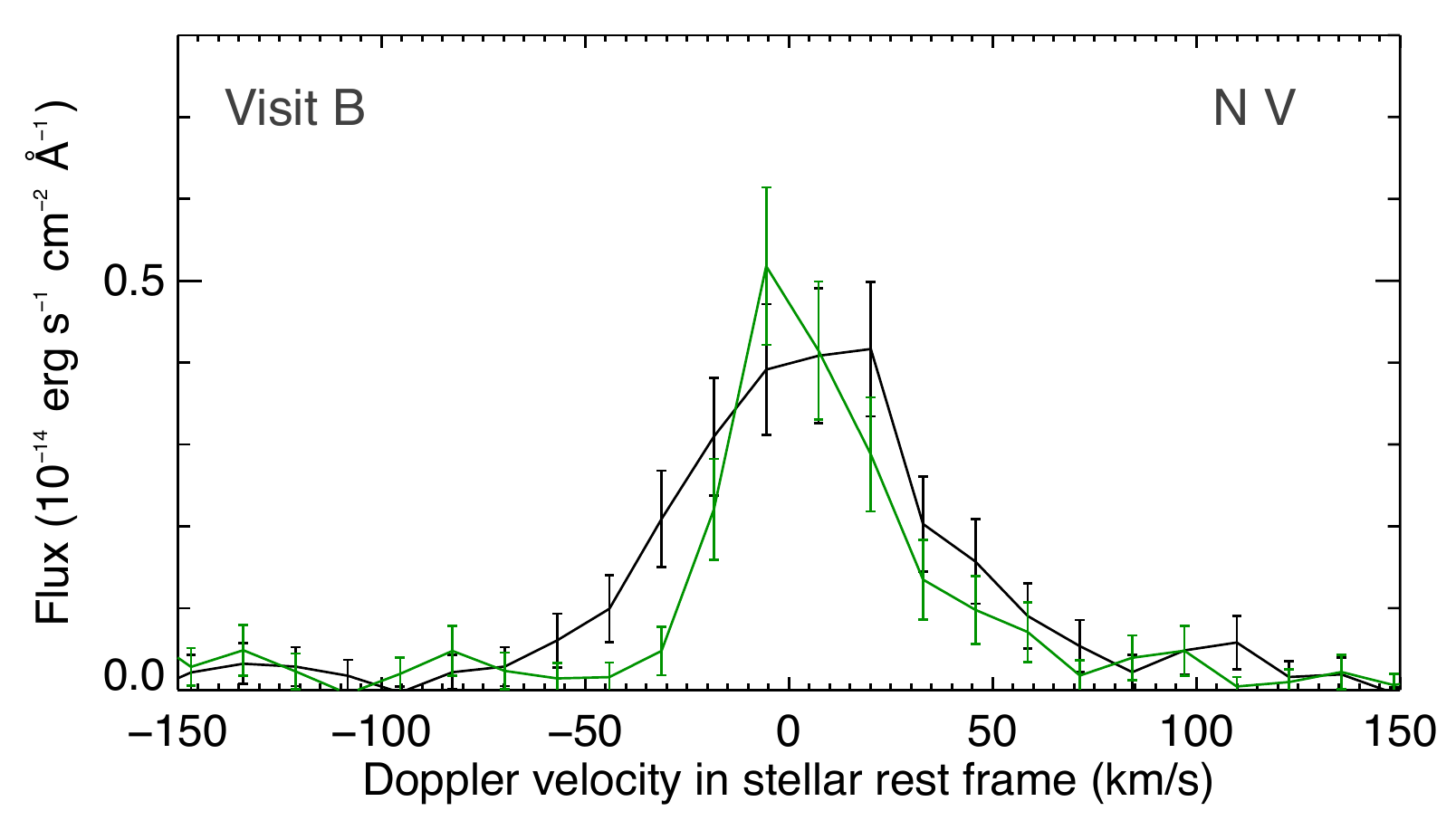}
\caption[]{\ion{N}{v}\,$\lambda$1243 line variations in Visit B. Same description as in Fig.~\ref{fig:Fig_SiIII_paper_VB}.}
\label{fig:Fig_NV2_paper_VB}
\end{figure}

No variations were detected in the brightest line of the \ion{N}{v} doublet (\,$\lambda$1239). The \ion{N}{v}\,$\lambda$1243 line, however, shows a lower flux in the pre-transit orbits as reported in \citealt{Bourrier2013}. The line is too faint to determine precisely the spectral ranges of these variations, but they appear to be located in the wings (Fig.~\ref{fig:Fig_NV2_paper_VB}). We consider orbits 3 and 4 to be most representative of the quiescent \ion{N}{v}\,$\lambda$1243 stellar line, as it is similar to that of quiescent lines in other visits, and its flux is consistent with half that in the \ion{N}{v}\,$\lambda$1239 line (0.47$\pm$0.04), as expected from the ratio of the lines oscillator strengths in an optically thin medium. In contrast the \ion{N}{v} lines flux ratio in orbits 1 and 2 is significantly lower than half (0.38$\pm$0.04).

\begin{figure}    
\includegraphics[trim=0cm 0cm 0cm 0cm,clip=true,width=0.99\columnwidth]{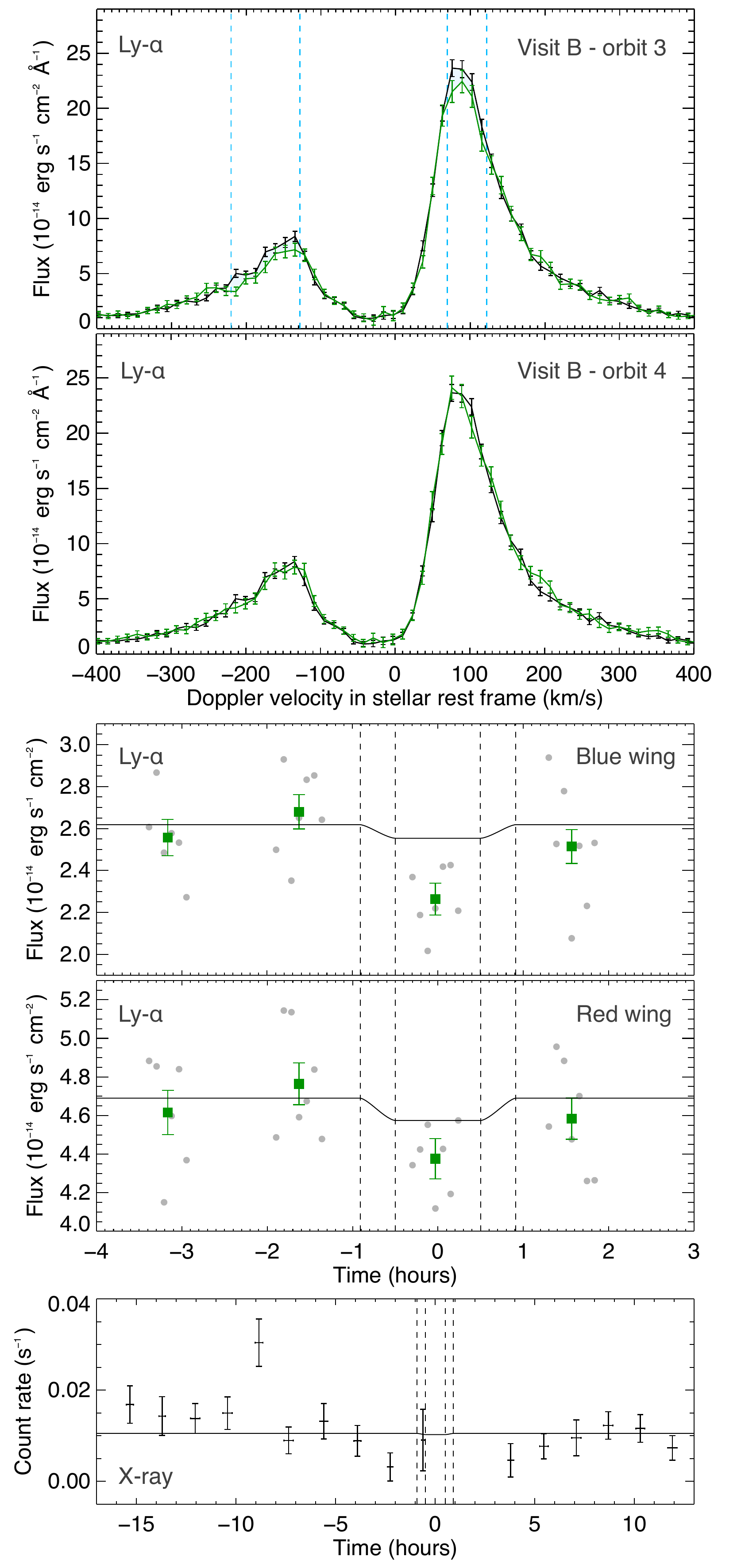}
\caption[]{\textbf{Top panels:} Stellar Lyman-$\alpha$ line in Visit B, before the optical transit of HD\,189733b (black spectrum), during (green spectrum, top subpanel), and after (green spectrum, bottom subpanel). \textbf{Middle panels:} flux integrated over the spectral ranges absorbed in orbit 3. \textbf{Bottom panel:} \textit{Swift}/XRT X-ray light curve obtained at the time of the transit in Visit B. Horizontal error bars correspond to the exposure time of data binned into one point per snapshot visit. The model transit light curve was normalised to the out-of-transit data, excluding the flare at about -9\,h.}
\label{fig:Fig_Ly_VB_paper}
\end{figure}

No variations are found in the Lyman-$\alpha$ line during the pre-transit orbits 1 and 2, which were taken as reference. We recover the significant absorption signature identified during in-transit orbit 3 by \citet{Lecav2012}, \citet{Bourrier2013}, with decrease in stellar flux by 14.1$\pm$3.6\% within -220.3 to -128.1\,km\,s$^{-1}$ (Fig.~\ref{fig:Fig_Ly_VB_paper}). Subtracting the 2.4\% absorption by the UV atmospheric continuum (assumed to be the same as measured in the optical by \citealt{Baluev2015}, see Sect~\ref{sec:cont_LC}) yields an excess absorption of 11.7$\pm$3.6\% by the exosphere of neutral hydrogen surrounding the planet. The flux decrease at the peak of the red wing in orbit 3 is marginal (6.7$\pm$2.7\% within 69.3 to 122.0\,km\,s$^{-1}$), even more so when correcting for the planetary continuum. While this decrease cannot be considered by itself as a clear signature of the planetary atmosphere (\citealt{Guo2016}), it nonetheless occurs at the same time as the significant flux decrease in the blue wing, and we discuss its possible planetary origin in light of the new visits in Sect.~\ref{sec:interp}. We do not detect any significant post-transit variation during orbit 4 (the blueshifted spectral range absorbed in orbit 3 yields a total variation of 5.2$\pm$3.9\%; see \citealt{Lecav2012} and \citealt{Bourrier2013}).\\

The temporal sampling of the \textit{Swift} light curve (Fig.~\ref{fig:Fig_Ly_VB_paper}) does not allow a comparison between the X-ray variations and those measured in the FUV, but HD\,189733 showed significant variability over the $\sim$28\,h of observations obtained before, during, and after the transit. The count rate decreased over the duration of visit B, and is interestingly lowest at the time of the planet transit. Furthermore a bright flare occurred about 8\,h before ingress, as previously noted by \citet{Lecav2012}. Unlike the flare in the \textit{XMM-Newton} data (see next visits), there was no centroid shift towards either the M star companion or the background source, showing that the flare in Visit B arose from the planet host star.\\


\subsection{Visit C - May 9/10, 2013}
\label{sec:VC}

\begin{figure}     
\includegraphics[trim=0cm 0cm 0cm 0cm,clip=true,width=\columnwidth]{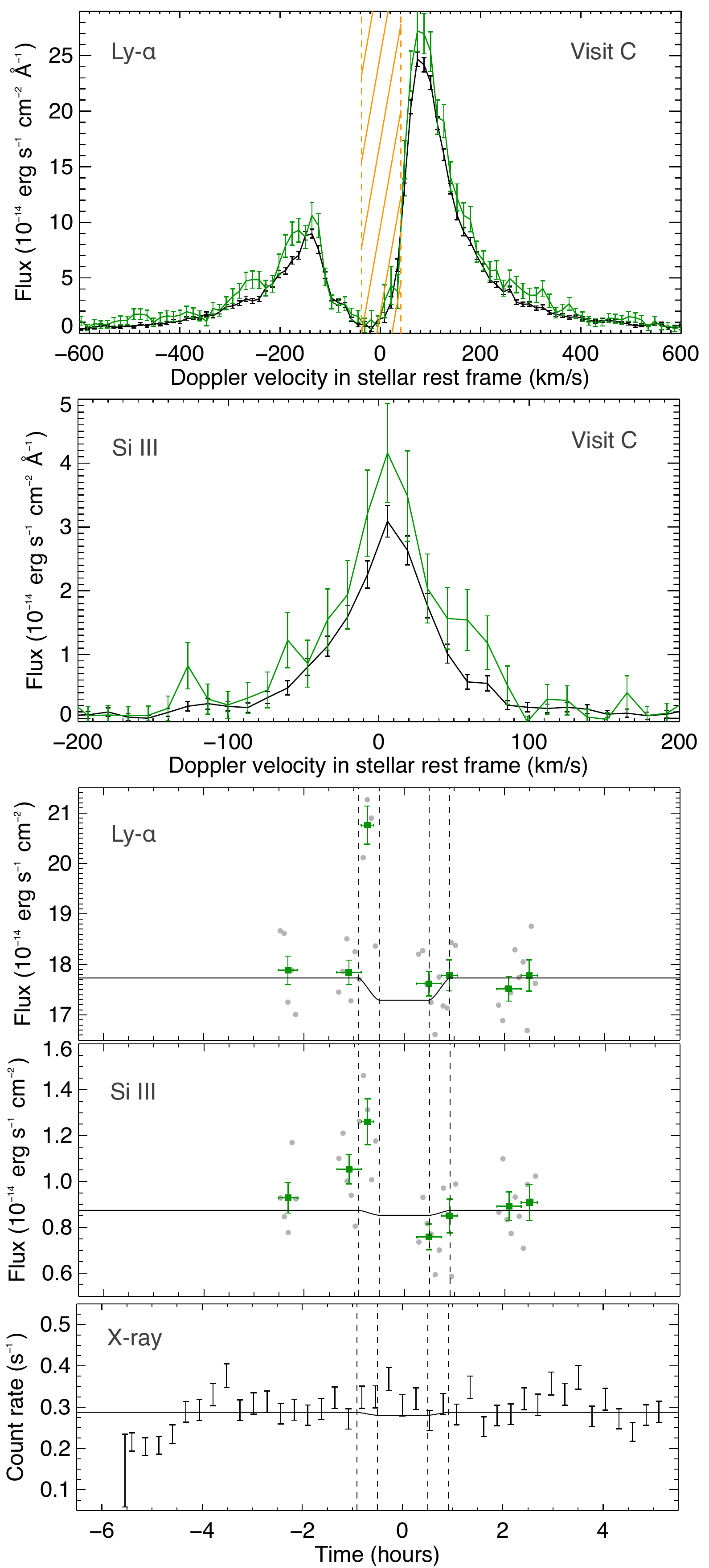}
\caption[]{FUV flare of HD\,189733 in Visit C. \textbf{Top panels} Lyman-$\alpha$ and \ion{Si}{iii} spectra displaying maximum flux increase during the flare (averaged over sub-exposures 7 to 9 in orbit 2, green profile), to be compared with the quiescent spectra (black profiles). \textbf{Middle panels} Temporal evolution of the flux in the flaring lines integrated over $\pm$400\,km\,s$^{-1}$. Green squares show sub-exposures (grey disks) binned by five (or by 6 and 3 in orbit 2) to better highlight the flare. \textbf{Bottom panel:} \textit{XMM-Newton} X-ray light curve in Visit D. There is no clear counterpart for the FUV flare at ingress.}
\label{fig:Fig_Ly_SiIII_VC_paper}
\end{figure}

In Visit C no significant variations were found in any of the lines between orbits 1, 3, and 4. A flare occurred in orbit 2 in the Lyman-$\alpha$ and \ion{Si}{iii} lines (Fig.~\ref{fig:Fig_Ly_SiIII_VC_paper}). The flux increase appears to be maximum in sub-exposures 7 to 9, during the ingress of HD\,189733b. It is possible that the flare began earlier in the \ion{Si}{iii} line, but the dispersion of the flux in individual sub-exposures makes it difficult to identify which ones exactly were affected. The flare increased the flux over the entire observed profiles of the Lyman-$\alpha$ and \ion{Si}{iii} lines, although the combination of ISM absorption and instrumental convolution prevents us from assessing whether the core of the intrinsic Lyman-$\alpha$ line was affected. Over the three most flaring sub-exposures, the \ion{Si}{iii} line increases by 44.9$\pm$12.5\% within $\pm$90\,km\,s$^{-1}$, and the Lyman-$\alpha$ line increases by 21.1$\pm$4.2\% over its blue wing and 16.3$\pm$2.7\% over its red wing (between the airglow boundaries and $\pm$400\,km\,s$^{-1}$). 

The Lyman-$\alpha$ and \ion{Si}{iii} lines do not appear redshifted during the flare, as is sometimes the case for chromospheric and transition region lines (e.g. \citealt{Pillitteri2015}, \citealt{Youngblood2017}). We see no sign of the flare in the \ion{N}{v} doublet, even though an increase on the order of that in the \ion{Si}{iii} line would have been detected. We also see no indication of the flare in the \ion{O}{v} and \ion{Fe}{xii} lines, although their faintness might prevent us from detecting such variations. The soft X-ray count rate is lower at the beginning of the visit but stabilises after about -3\,h, and shows no evidence for the flare (Fig.~\ref{fig:Fig_Ly_SiIII_VC_paper}). \\

FUV-only flares have previously been observed in G-type stars and M dwarfs (\citealt{MitraKraev2005}, \citealt{Ayres2015}, \citealt{Loyd2018_MUSCLESV}). Flares result from reconnections occurring in magnetic structures. These reconnections accelerate electrons along the reconnecting field lines. When the electron beam impacts the dense lower atmosphere of the star, it rapidly heats the gas which expands to fill the reconnecting loop, producing the soft X-ray emission of the flare. The total energy available for this depends on the free magnetic energy available in the reconnecting loop. Loops with a greater reservoir of free energy can produce more energetic, higher temperature flares. \citet{Loyd2018_MUSCLESV} suggest that UV-only flares may be the result of reconnections in smaller magnetic structures that are only capable of heating the stellar transition region and chromosphere, whereas larger structures may release enough energy to drive hotter X-ray emitting plasma up into the corona. This scenario would be consistent with the relatively low amplification factors measured in HD\,189733 Lyman-$\alpha$ and \ion{Si}{iii} lines, and the non-detection of the flare in the higher-energy lines and X-rays. The stronger amplification in the \ion{Si}{iii} line suggests that the energy released by the flare peaks at lower temperatures. 

The observed flare is characterised by an abrupt rise in the Lyman-$\alpha$ line and a longer rise in the \ion{Si}{iii} line (Fig.~\ref{fig:Fig_Ly_SiIII_VC_paper}). While our observations do not cover the decay phase, the flux in both lines appear to return to its quiescent level in a short time ($\sim$50\,min at maximum). This behaviour is consistent with that of the two flares observed by \citet{Pillitteri2015} in the FUV, which had short duration of 1\,h and 400\,s maximum, and also showed a larger flux increase in the \ion{Si}{iii} line compared to the \ion{N}{v} doublet. On the other hand, lower optical chromospheric lines observed during a flare of HD\,189733 with UVES showed a long decay after the initial short rise phase (\citealt{Czesla2015}, \citealt{Klocova2017}). These differences between optical and FUV lines likely traces a different behaviour between the lower and upper chromosphere of HD\,189733 during flares.

\subsection{Visit D - November 3, 2013}
\label{sec:VD}

\begin{figure}    
\includegraphics[trim=0cm 0cm 0cm 0cm,clip=true,width=\columnwidth]{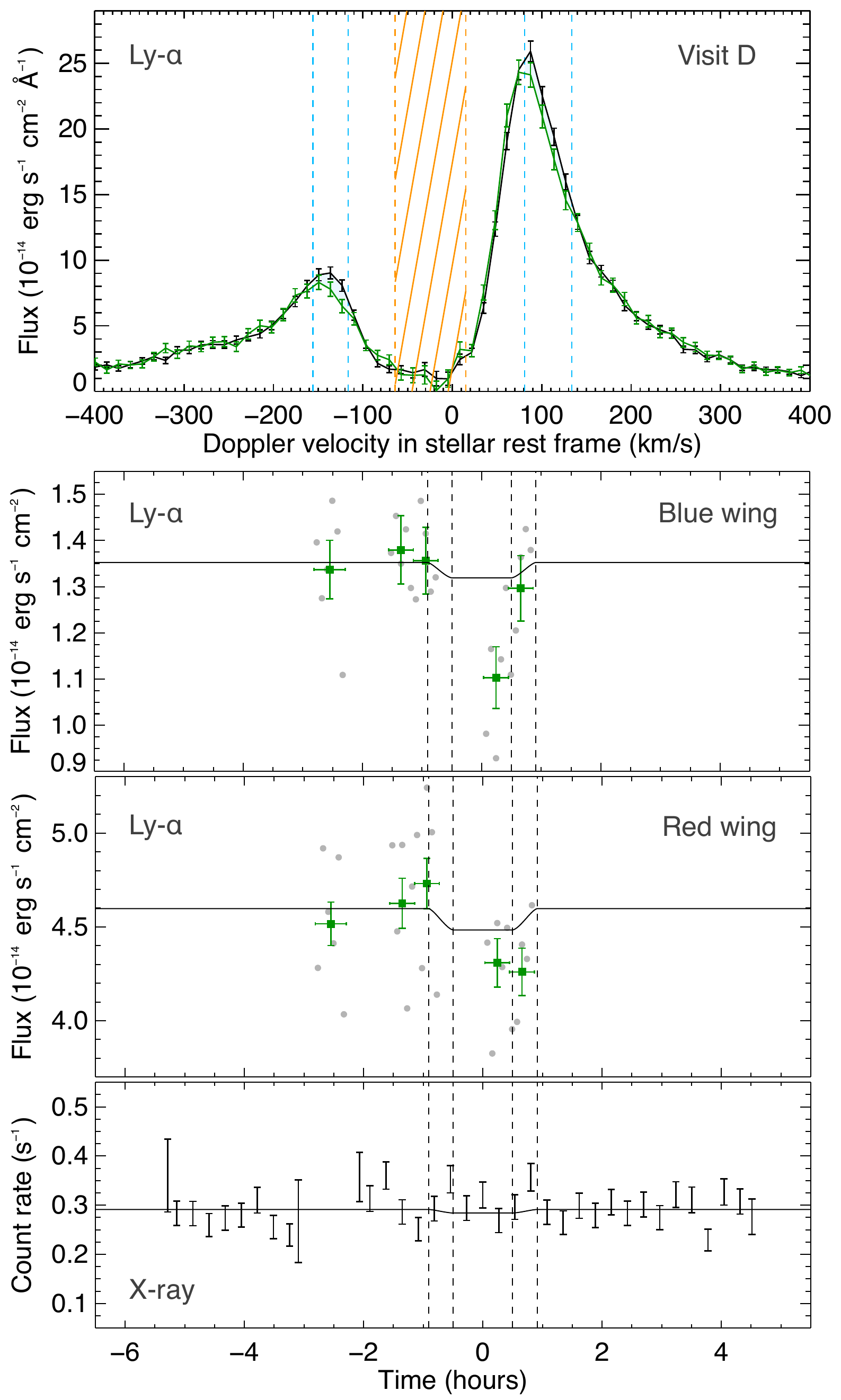}
\caption[]{\textbf{Top panel:} Stellar Lyman-$\alpha$ line in Visit D. Spectral ranges showing flux decreases in orbit 3 (green spectrum) compared to pre-transit orbits (black spectrum) as highlighted in blue. \textbf{Middle panels:} flux integrated over the spectral range absorbed in orbit 3. Grey disks correspond to sub-exposures, binned by 6 or 5 (green squares). \textbf{Bottom panel:} \textit{XMM-Newton} X-ray light curve obtained at the time of the planet transit in Visit D. The model transit light curve was normalised to the out-of-transit data. Data contaminated by a flare from the M dwarf companion at about -3\,h has been excluded.}
\label{fig:Fig_Ly_VD_paper}
\end{figure}

In Visit D no significant variations were found in any of the lines between orbits 1 and 2. In orbit 3 the Lyman-$\alpha$ line shows two significant flux decreases (Fig.~\ref{fig:Fig_Ly_VD_paper}), in the blue wing (13.2$\pm$4.4\% between -156.1 and -116.5\,km\,s$^{-1}$) and in the red wing (7.6$\pm$2.5\% between 80.8 and 133.6\,km\,s$^{-1}$). The flux does not vary significantly over the region in between these signatures (-5.7$\pm$3.0\% between -116.5 and 80.8\,km\,s$^{-1}$ minus the airglow). These absorption signatures are reminiscent of those detected in Visit B: they occur during the planetary transit, have consistent absorption depths, and are located within similar spectral regions (although the blue wing signature in Visit D is less blueshifted than in Visit B). While the Visit D red wing signature remains stable during transit, the blue wing signature is deeper at mid-transit than its counterpart in Visit B (21.3$\pm$5.7\% in the first half of orbit 3) and disappears during egress (Fig.~\ref{fig:Fig_Ly_VD_paper}).

\begin{figure*}     
\begin{minipage}[tbh]{\textwidth}
\includegraphics[trim=0cm 0cm 0cm 0cm,clip=true,width=\columnwidth]{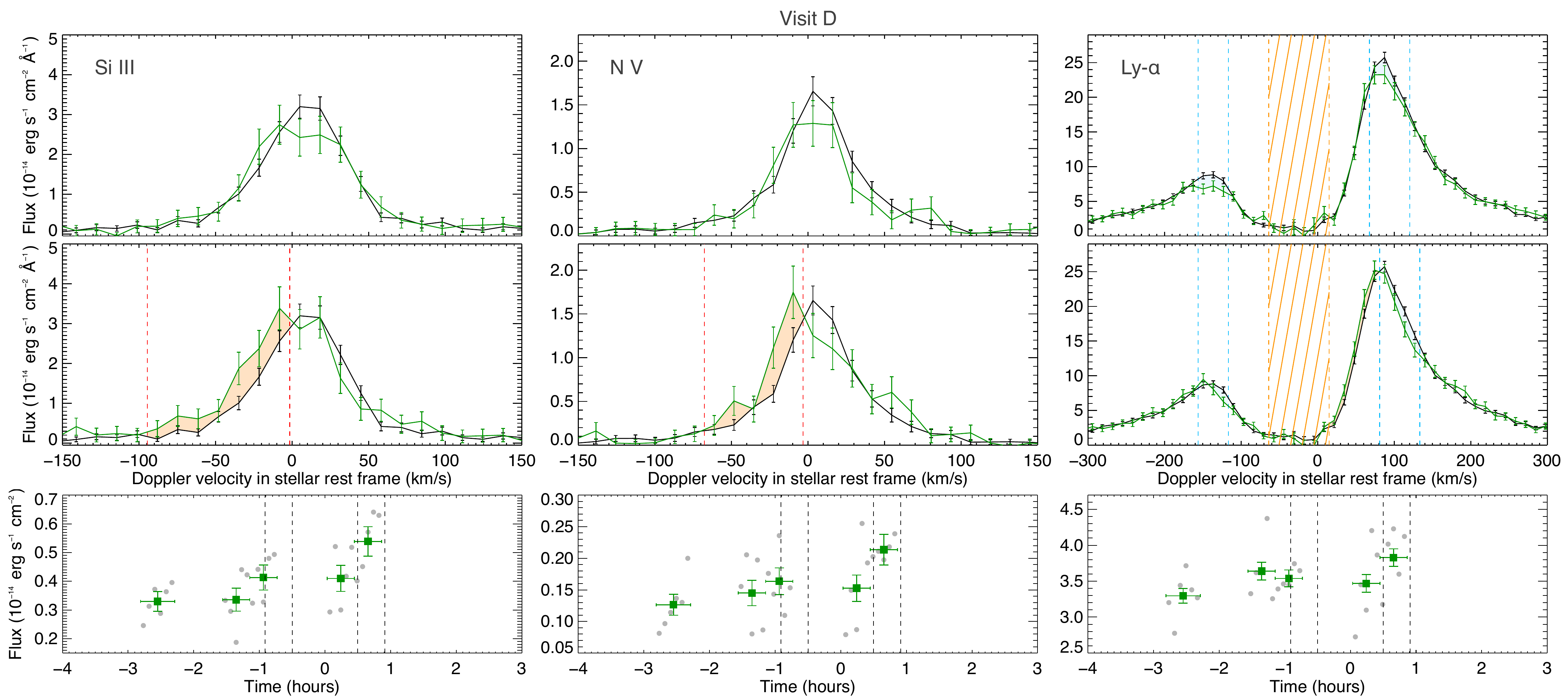}
\caption[]{Variations in the \ion{Si}{iii} line (left column), \ion{N}{v} doublet (middle column), and Lyman-$\alpha$ line (right column) in Visit D. Green spectra show the lines during the first half (top row) and the second half (middle row) of orbit 3. Their comparison with the quiescent stellar lines (black spectra) reveals simultaneous flux increases during egress (highlighted as red regions). Blue regions highlight the absorbed spectral ranges in the Lyman-$\alpha$ line. Bottom panels show the temporal evolution of the flux integrated over the flaring spectral regions. Green squares show sub-exposures (grey disks) binned by 6 or 5.}
\label{fig:Fig_SiIII_NV_Ly_VD_paper}
\end{minipage}
\end{figure*}

We caution that the red wing signature has only marginal transit depth when accounting for the planetary continuum. Furthermore, we found that the Lyman-$\alpha$ flux in between the airglow and the red wing signature in fact increases sharply at egress during the second half of orbit 3 (by 16.0$\pm$5.5\% within 15.0 to 67.8\,km\,s$^{-1}$). A simultaneous increase in flux occurs in the blue wings of the \ion{Si}{iii} line (53.0$\pm$17.5\%) and \ion{N}{v} doublet (52.2$\pm$21.0\%), as can be seen in Fig.~\ref{fig:Fig_SiIII_NV_Ly_VD_paper}. These variations might be correlated and trace the onset of a flare in the stellar chromosphere, making it difficult to determine the exact properties of the Lyman-$\alpha$ transit signatures. 

There are no significant variations in the soft X-rays emitted by HD\,189733 during Visit D, in particular during egress (Fig.~\ref{fig:Fig_SiIII_NV_Ly_VD_paper}), although the example of Visit C shows that flares from this star can be limited to FUV lines. \\

\begin{figure}    
\centering
\includegraphics[trim=1cm 1.5cm 11.5cm 19.cm,clip=true,width=0.8\columnwidth]{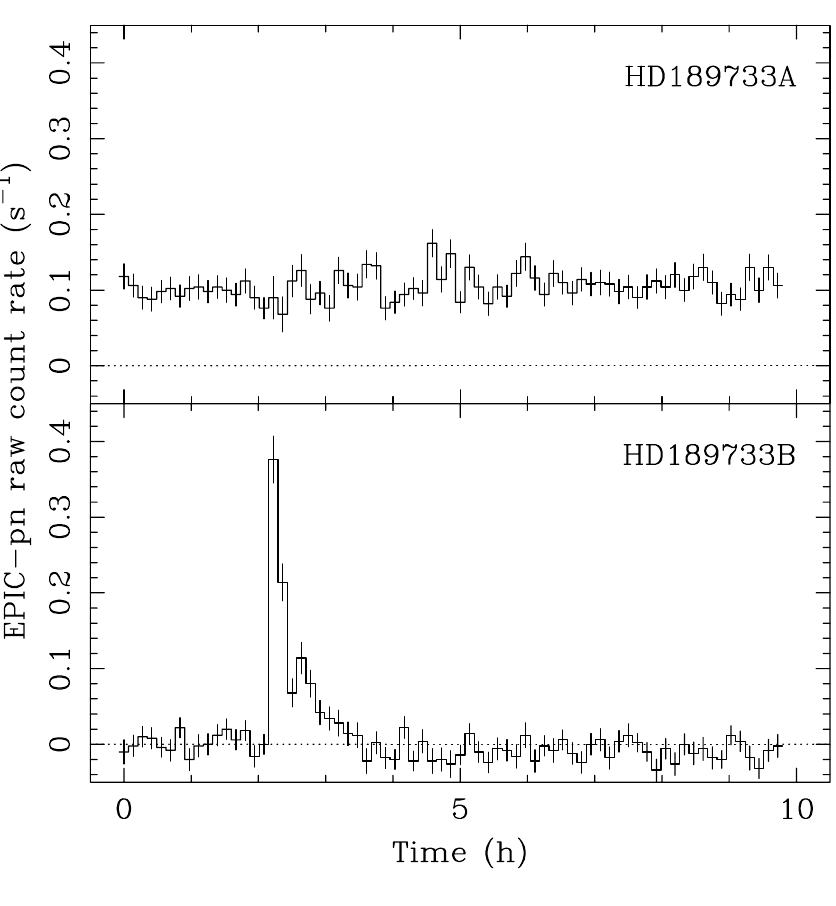}
\caption[]{Comparison of the count rates for HD\,189733 and its M star companion HD189733B in Visit D. Time is relative to the start of the X-ray observations. This clearly shows the flare is from the companion star.}
\label{fig:VisitD_Xray_comp}
\end{figure}

We note that a strong flare did occur in the X-ray light curve about 3\,h before mid-transit. Fig.~\ref{fig:VisitD_Xray_comp} shows the separate X-ray count rates from each component of the system separately, obtained using the method described in Sect.~\ref{sec:xray_obs}. This clearly shows the flare to have been from the M dwarf companion HD\,189733B, which calls in question the conclusion by \citet{Poppenhaeger2013} that it is inactive. Positional analysis confirms that the M dwarf is the origin of the flare, as the X-ray centroid is seen to shift to its position during the flare. Fig.~\ref{fig:regions} highlights this, where the 0.15 -- 1.0\,keV image is dominated by emission from the position of primary star (leftmost panel) at all times except during the flare, when it shifts to the companion (second panel from the left). The flare is also seen originating from the M dwarf in the ultraviolet with the XMM OM (fourth panel of Fig.~\ref{fig:regions}). The time period of the flare, which was covered by the first orbit of the \textit{HST} visit, has been excluded in Fig.~\ref{fig:Fig_SiIII_NV_Ly_VD_paper} and from the rest of the analysis. We do not see any evidence for the flare in \textit{STIS} 2D images. We note that even in a case where HD\,189733B would enter the 52x0.1 arcsecond-wide slit used in this visit, its spectrum would be located about 420 pixels from the spectrum of HD\,189733 and would thus not contaminate its extraction. The variations in FUV lines discussed above thus arise from the primary HD\,189733 or its planetary companion.

\subsection{Visit E - November 21, 2013}
\label{sec:VE}

\begin{figure}    
\includegraphics[trim=0cm 0cm 0cm 0cm,clip=true,width=\columnwidth]{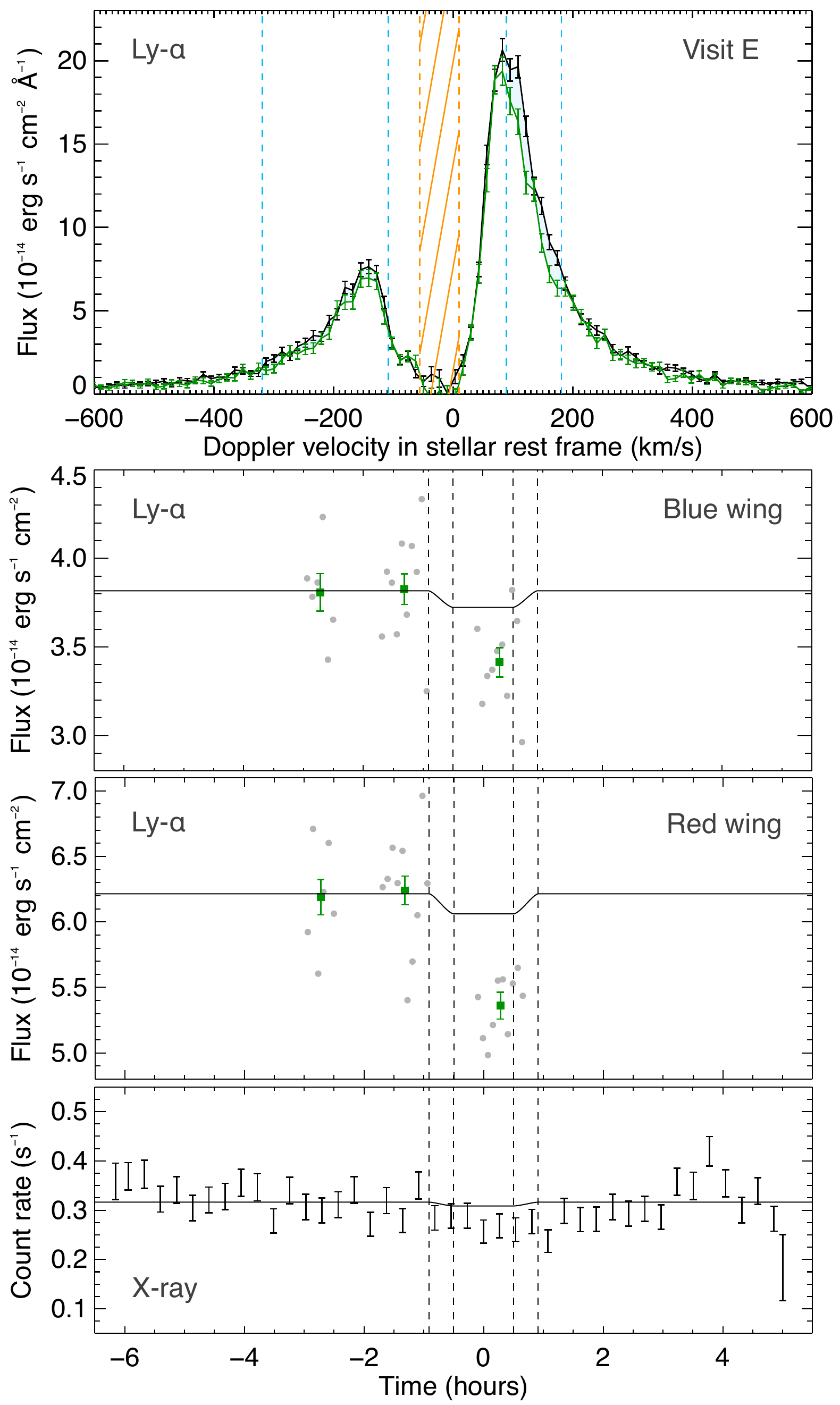}
\caption[]{\textbf{Top panel:} Stellar Lyman-$\alpha$ line in Visit E. Spectral ranges showing the clearest flux decreases in orbit 3 (green spectrum) compared to pre-transit orbits (black spectrum) as highlighted in blue. \textbf{Middle panels:} flux integrated over the spectral ranges shown in the top panel. Grey disks correspond to sub-exposures, green squares to orbit-averaged exposures. \textbf{Bottom panel:} \textit{XMM-Newton} X-ray light curve obtained at the time of the planet transit in Visit E. The model transit light curve was normalised to the out-of-transit data.}
\label{fig:Fig_Ly_VE_paper}
\end{figure}

The Lyman-$\alpha$ line shows no significant variations in orbits 1 and 2. In orbit 3 both wings of the line show clear signatures, decreasing by 13.7$\pm$2.0\% within 75.8 and 181.2\,km\,s$^{-1}$, and by 10.6$\pm$2.7\% within -319.0 and -108.3\,km\,s$^{-1}$ (Fig.~\ref{fig:Fig_Ly_VE_paper}). Strong flux decreases are also measured at larger velocities in both wings (21.6$\pm$6.8\% within -503.3 and -358.4\,km\,s$^{-1}$; 33.1$\pm$7.7\% within 352.1 and 404.9\,km\,s$^{-1}$; 46.2$\pm$8.2\% within 510.0 and 589.1\,km\,s$^{-1}$). Even though the observed Lyman-$\alpha$ line does not trace directly the intrinsic stellar line because of the combination of ISM absorption and instrumental convolution, we note that the decrease is on the same order ($\sim$12\%) for the main signatures in the blue and red wings of the line, and that the core of the observed line remains stable (-2.6$\pm$3.5\% within -108.3 and 75.8\,km\,s$^{-1}$, airglow excluded).

Interestingly the soft X-ray count rate drops in the same time window as the Lyman-$\alpha$ flux, before increasing suddenly about 3\,h after mid-transit (Fig.~\ref{fig:Fig_Ly_VE_paper}). This tentatively suggests that the FUV flux decrease has a counterpart in the X-ray (Sect.~\ref{sec:interp}).

No significant variations are detected in the \ion{Si}{iii} and the bright \ion{N}{v}\,$\lambda$1239 lines. The \ion{N}{v}\,$\lambda$1243 line shows a similar flux between orbits 2 and 3, but is nearly twice brighter in orbit 1 (by 81.2$\pm$24.8\%). Surprisingly the \ion{N}{v}\,$\lambda$1243 flux in orbit 1 and 2+3 are respectively 0.670$\pm$0.079 and 0.370$\pm$0.036 times that of the average flux in the \ion{N}{v}\,$\lambda$1239 line, suggesting that none of the exposures is representative of the quiescent stellar line. This is further supported by the reconstruction performed in Sect.~\ref{sec:EUV_spec}, which showed that the \ion{N}{v}\,$\lambda$1243 line in either orbit 1 or orbit 2+3 is not well fitted with the properties derived for the \ion{N}{v}\,$\lambda$1239 intrinsic line. While we do not know the origin of the large flux variation in the \ion{N}{v}\,$\lambda$1243 line, its average over orbits 1, 2, and 3 likely best represents the quiescent stellar line, as it is about half as bright as the \ion{N}{v}\,$\lambda$1239 line (0.47$\pm$0.04) and is well fitted with its derived properties.

\subsection{Continuum light curves}
\label{sec:cont_LC}

In previous sections, we identified spectra displaying strong flux variations in emission lines that could be caused by the star or by clouds of atoms and ions surrounding the planet. Other spectra showing no detectable variations were considered as representative of the quiescent stellar lines. However, during the transit these spectra are still absorbed by the planetary UV atmospheric continuum, which we sought to measure. We assumed a grey absorber yielding the same opacity at all wavelengths within a given stellar line, but varying with the spectral region. We thus fitted independent transit light curves to the flux integrated over the Lyman-$\alpha$, \ion{Si}{iii}, and the cumulated \ion{N}{v} lines, using the \textsc{exofast} routines (\citealt{Mandel2002}, \citealt{Eastman2013}). Given the precision and temporal coverage of our data we assumed a uniform stellar disk with no limb-darkening or brightening. Visit E was not included in the fit to the Lyman-$\alpha$ and \ion{N}{v} light curves, as its in-transit exposures show strong variability in these lines (Sect.~\ref{sec:VE}). We fitted the planet-to-star radii ratio $R_\mathrm{p}^\mathrm{line}/R_\mathrm{*}$, and fixed all other system properties to the values given in Table~\ref{tab:param_sys}. \\  

\begin{figure}    
\includegraphics[trim=0cm 0cm 0cm 0cm,clip=true,width=\columnwidth]{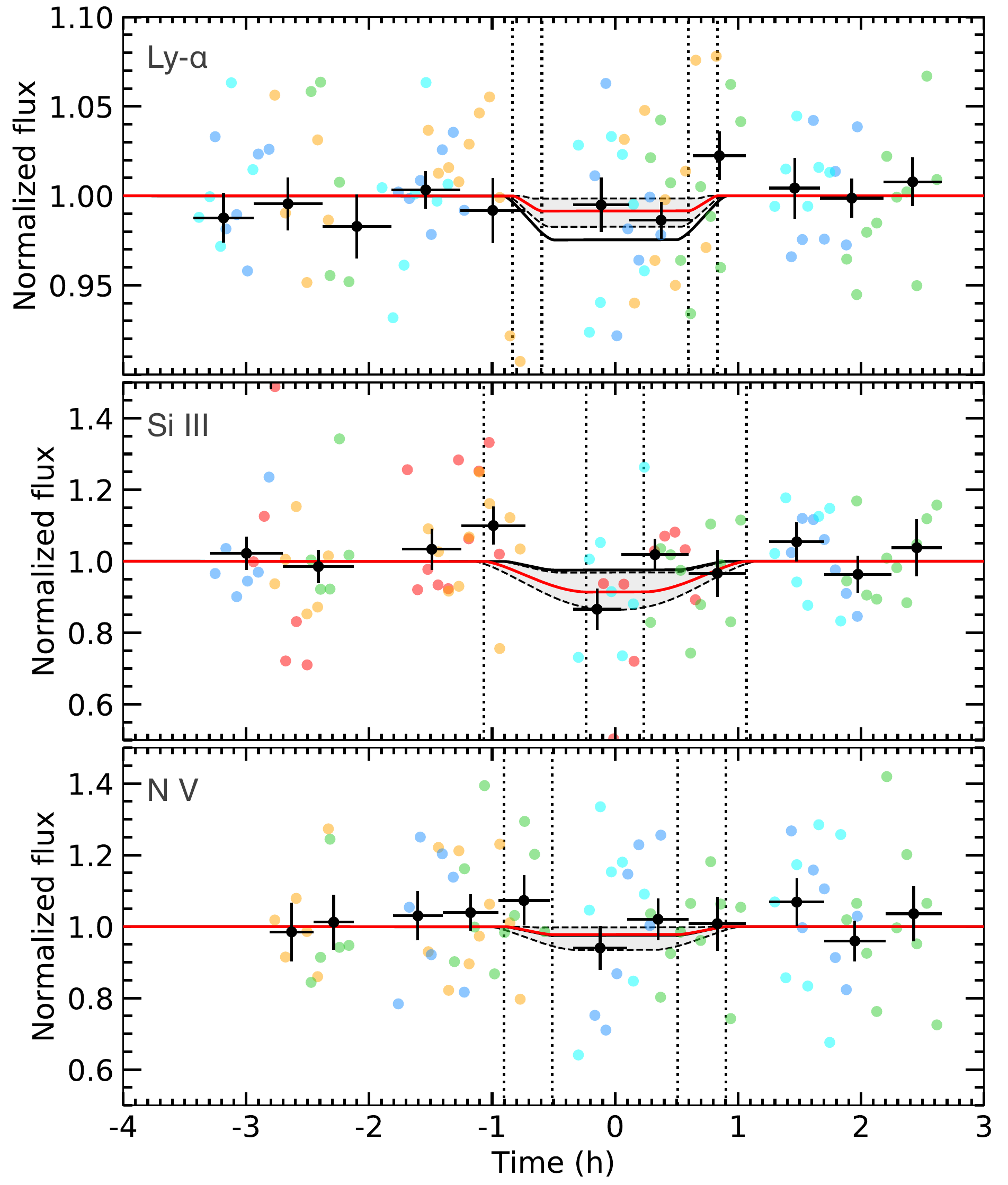}
\caption[]{Transit light curves of HD\,189733b in the Lyman-$\alpha$ (top), \ion{Si}{iii} (middle), \ion{N}{v} (bottom) lines. The flux has been integrated over the largest spectral range showing no strong variations in each visit, and only visits with stable sub-exposures both out- and in- transit have been fitted with a transit model. The best-fit is shown as a red line, with the corresponding contact times highlighted with dotted vertical lines. Gray bands delimit the 1$\sigma$ envelopes of the best-fit models. The solid black line shows the transit light curve with the planet optical radius in front of a uniform disk, for comparison. Sub-exposures are coloured in deep blue for Visit A, cyan for Visit B, green for Visit C, orange for Visit D, and red for Visit E. Black points show binned sub-exposures.}
\label{fig:Fig_LC}
\end{figure}

We sampled the posterior distributions of the model parameters using the Markov-Chain Monte Carlo (MCMC) Python software package \textsc{emcee} (\citealt{Foreman2013}). The best-fit transit light curves in the region of each line, shown in Fig.~\ref{fig:Fig_LC}, correspond to $R_\mathrm{p}^\mathrm{Ly-\alpha}/R_{\star}$ = 0.092$\stackrel{+0.039}{_{-0.053}}$, $R_\mathrm{p}^\mathrm{\ion{Si}{iii}}/R_{\star}$ = 0.29$\stackrel{+0.08}{_{-0.12}}$, $R_\mathrm{p}^\mathrm{\ion{N}{v}}/R_{\star}$ = 0.15$\pm$0.10. These values reveal a marginal transit detection in the \ion{Si}{iii} line (2.4\,$\sigma$), but are not significantly different from 0 ($<$3\,$\sigma$) and consistent with the optical transit ($R_\mathrm{p}/R_{\star}$ = 0.1571$\pm$0.0004, corresponding to a transit depth of 2.4\%, \citealt{Baluev2015}). Fitting a common transit model to the combined Lyman-$\alpha$, \ion{Si}{iii}, and \ion{N}{v} fluxes yield $R_\mathrm{p}^\mathrm{FUV}/R_{\star}$ = 0.104$\stackrel{+0.036}{_{-0.049}}$, similarly consistent with the optical transit and different from 0 by only 2\,$\sigma$. The present data are therefore not precise enough to measure the atmospheric continuum of HD\,189733b in the FUV.   \\

\section{Analysis of the high-energy stellar spectrum}
\label{sec:HE_spec}

We used the quiescent X-ray and FUV spectra identified in Sect.~\ref{sec:search_var} to analyse the high-energy spectrum of HD\,189733 and its evolution over the different observing epochs. Spectra obtained during the planetary transit were corrected for the planetary continuum absorption, fixed to 2.4\% (see Sect~\ref{sec:cont_LC}).

\subsection{Retrieval of the stellar Ly-$\alpha$ line and ISM properties.}
\label{sec:Ly_fit}

Knowledge of the intrinsic stellar Lyman-$\alpha$ line profile is important to our understanding of the stellar chromosphere, its extreme ultraviolet (EUV) emission, and its impact on the planetary atmosphere. We reconstructed the theoretical line profiles in each epoch, following the same procedure as in, e.g., \citet{Bourrier2015_GJ436,Bourrier2017_HD976}. A model profile of the intrinsic stellar line is absorbed by hydrogen and deuterium in the ISM, convolved with \textit{STIS} line spread function (LSF, as derived by \citealt{Bourrier2017_HD976}), and fit to the quiescent Lyman-$\alpha$ lines in each visit. These master spectra were built by co-adding the flux over the spectral ranges identified as stable in each exposure. The model is oversampled in wavelength, and rebinned over the \textit{STIS} spectral table after convolution. The comparison between observed and theoretical data was performed between -400 and 400\,km\,s$^{-1}$ (defined in the star rest frame), excluding the airglow-contaminated spectral ranges (Fig.~\ref{fig:Fig_airglow}). We sampled the posterior distributions of the master spectra parameters using \textsc{emcee}, and derived their best-fit values and uncertainties using the same method as in \citet{Bourrier2018_GJ3470b}.\\

We performed preliminary simulations to determine the best model for the intrinsic stellar line, using the BIC as merit function. This revealed three interesting features: (i) the intrinsic line is best fitted with a double-peaked Voigt profile (compared to other typical line profiles with a single peak or Gaussian components, e.g. \citealt{Wood2005}, \citealt{Youngblood2016}), confirming previous findings by \citet{Bourrier2013}; (ii) there is no significant difference in the velocity separation between the two peaks (assumed to be identical) between the different visits; (iii) in each visit, the intrinsic stellar line is well centred on the Lyman-$\alpha$ transition wavelength in the star rest frame. The final theoretical intrinsic line was thus modelled as two Voigt profiles with the same total flux, temperature (assuming pure thermal broadening), and damping parameter. These three free parameters are specific to each visit. The two Voigt profiles are placed at the same distance from the Lyman-$\alpha$ transition wavelength in the star rest frame, and their velocity separation is a free parameter common to all visits.\\

The theoretical absorption profile of the ISM along the line of sight is common to all visits, and defined by its column density of neutral hydrogen log$_{10}$\,$N_{\mathrm{ISM}}$(H\,{\sc i}), its temperature $T_{\mathrm{ISM}}$ and turbulent velocity $\xi_{\mathrm{ISM}}$, and its heliocentric radial velocity $\gamma_{\mathrm{ISM}/\sun}$. The D\,{\sc i}/H\,{\sc i} ratio was set to 1.5$\times$10$^{-5}$\ ({e.g.,} \citealt{Wood2004}, \citealt{Hebrard_Moos2003}; \citealt{Linsky2006}). The spectral resolution of the \textit{STIS} data and the small difference in mass between hydrogen and deuterium prevent us from constraining both the temperature and turbulent velocity, and the latter was thus fixed to a constant value. The LISM kinematic calculator\footnote{\mbox{\url{http://sredfield.web.wesleyan.edu/}}}(\citealt{Redfield_Linsky2008}) predicts that the line of sight (LOS) toward HD\,189733 crosses the Mic ($T_{\mathrm{Mic}}$=9900$\pm$2000\,K, $\xi_{\mathrm{Mic}}$=3.1$\pm$1.0\,km\,s$^{-1}$, $\gamma_{\mathrm{Mic}/\sun}$=-22.2$\pm$1.3\,km\,s$^{-1}$), Eri ($T_{\mathrm{Eri}}$=5300$\pm$4000\,K, $\xi_{\mathrm{Eri}}$=3.6$\pm$1.0\,km\,s$^{-1}$, $\gamma_{\mathrm{Eri}/\sun}$=-15.9$\pm$1.0\,km\,s$^{-1}$), and Aql ($T_{\mathrm{Aql}}$=7000$\pm$2800\,K, $\xi_{\mathrm{Aql}}$=2.1$\pm$0.6\,km\,s$^{-1}$, $\gamma_{\mathrm{Aql}/\sun}$=-18.6$\pm$1.0\,km\,s$^{-1}$) clouds. In a first step, we assumed that a single cloud contributes to the ISM opacity along the LOS, and we fixed $\xi_{\mathrm{ISM}}$ to the error-weighted mean of the three clouds turbulent velocities. This led to an ISM cloud with $\gamma_{\mathrm{ISM}/\sun}$ = -21.3$\pm$0.5\,km\,s$^{-1}$, which suggests that the Mic cloud is the dominant ISM opacity source. However, the single-cloud model yields $T_{\mathrm{ISM}}$ = 15365$\pm$500\,K, which is much larger than temperatures expected for the local ISM (\citealt{Redfield_Linsky2008}), in particular in the direction of HD\,189733. We thus performed the final fit using a two-cloud ISM model. The turbulent velocity of component A was fixed to that of the MIC cloud, and the turbulent velocity of component B was fixed to the error-weighted mean of the Eri and Aql cloud values. Other properties were let free to vary independently for each component. \\

\begin{figure*}
\centering
\includegraphics[trim=0cm 0cm 0cm 0cm,clip=true,width=\textwidth]{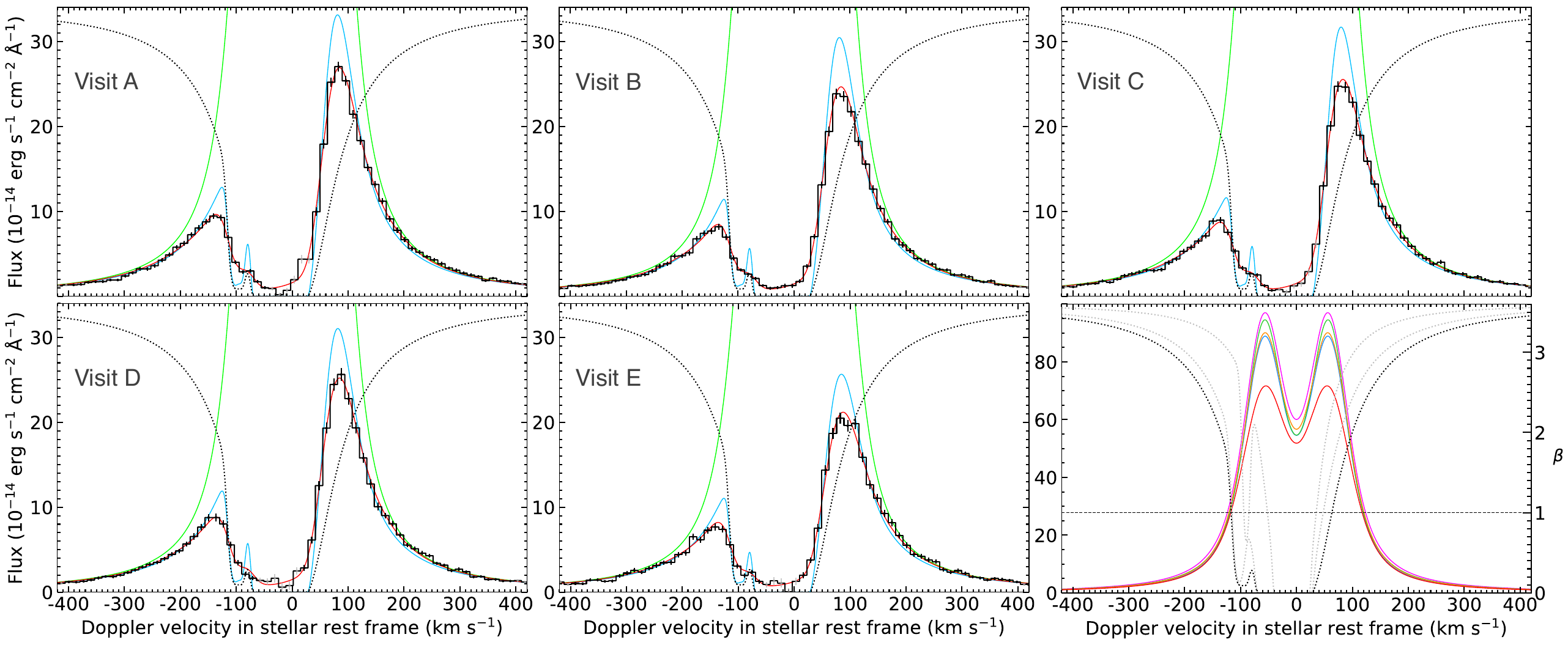}
\caption[]{Quiescent Lyman-$\alpha$ line profiles of HD\,189733. Black histograms show observed spectra, fit over points with black error bars. Green lines are the best fits for the intrinsic stellar line profiles in each visit, at Earth distance. They yield the blue profile after absorption by the interstellar medium, and the red profile after further convolution with the \textit{STIS} LSF. The bottom right panel shows a comparison of the intrinsic stellar line profiles between the visits (coloured in purple, blue, green, orange, and red from Visit A to E), with the corresponding radiation pressure to stellar gravity $\beta$ ratio reported on the right axis. The total ISM absorption profile is plotted as a dotted black line in all panels (scaled to the panel vertical range), with the absorption from individual ISM components plotted shown as dotted grey lines in the bottom right panel.}
\label{fig:La_stellar_lines}
\end{figure*}

Best-fit models are shown in Fig.~\ref{fig:La_stellar_lines}. They yield a good $\chi^2$ of 277 for 260 degrees of freedom ($\chi^2_\mathrm{r}$ = 1.07, with 282 datapoints and 22 free parameters). Properties of interest for the Lyman-$\alpha$ line and ISM are given in Table~\ref{table:tab_paramsfit}. The double-cloud ISM model improves the $\chi^2$ with no change in the BIC compared to the single-cloud ISM model ($\chi^2$ = 292, BIC = 400). The properties of component A ($T_{\mathrm{A}}$=13074$\stackrel{+1137}{_{-1083}}$\,K, $\gamma_{\mathrm{A}/\sun}$=-23.7$\stackrel{+1.2}{_{-1.7}}$\,km\,s$^{-1}$) are more consistent with those of the Mic cloud, while component B has a temperature in between those of the Eri and Aql clouds but is more redshifted ($T_{\mathrm{B}}$=5755$\stackrel{+2910}{_{-2802}}$\,K, $\gamma_{\mathrm{B}/\sun}$=-10.8$\stackrel{+2.2}{_{-2.4}}$\,km\,s$^{-1}$). The column densities of both components (Table~\ref{table:tab_paramsfit}) are in the range expected for a star at a distance of 19.8\,pc (Fig. 14 in \citealt{Wood2005}).

\begin{table*}
\caption{Properties derived from the fit to the HD\,189733 stellar lines and the reconstructed XUV spectrum.}
\centering
\begin{threeparttable}
\begin{tabular}{llccccccl}
\hline
\noalign{\smallskip}  
    &   \textbf{Parameter}       & \textbf{Visit A}      & \textbf{Visit B}         & \textbf{Visit C}      & \textbf{Visit D}      & \textbf{Visit E}  & \textbf{Average}      & \textbf{Unit} \\                           
\noalign{\smallskip}
\hline
\hline
\noalign{\smallskip}
\textbf{Stellar lines} &      &  &   &   &   &       &      &   \\
\noalign{\smallskip}
Lyman-$\alpha$ &  $F$    & 14.24$\stackrel{+0.63}{_{-0.59}}$ & 12.83$\stackrel{+0.59}{_{-0.54}}$  & 13.38$\stackrel{+0.63}{_{-0.60}}$  & 13.21$\stackrel{+0.59}{_{-0.60}}$  & 11.25$\stackrel{+0.51}{_{-0.43}}$      &   12.77$\pm$0.25     &  \,erg\,cm$^{-2}$\,s$^{-1}$ \\
\ion{Si}{ii} &  $\gamma$    & \multicolumn{6}{c}{6.6$\stackrel{+6.6}{_{-5.3}}$}     & km\,s$^{-1}$  \\
               &  $F$     & \multicolumn{6}{c}{2.67$\stackrel{+0.60}{_{-0.53}}$} & 10$^{-3}$\,erg\,cm$^{-2}$\,s$^{-1}$  \\
               &  $T$    & \multicolumn{6}{c}{9.6$\stackrel{+14.6}{_{-6.9}}$} & 10$^{5}$\,K  \\
\ion{Si}{iii} &  $\gamma$     & 7.9$\pm$1.2 & 6.2$\pm$1.0  & 6.3$\pm$1.3  & 7.4$\pm$1.4  & 5.8$\pm$1.0      &  6.77$\pm$0.53     & km\,s$^{-1}$  \\
               &  $F$    & 185.6$\stackrel{+7.8}{_{-6.9}}$  & 176.6$\pm$7.4  & 157.1$\pm$7.6  & 164.6$\pm$7.4  & 149.6$\pm$6.0 & 166.5$\stackrel{+3.6}{_{-3.3}}$     & 10$^{-3}$\,erg\,cm$^{-2}$\,s$^{-1}$  \\
               &  $T$    & 12.0$\pm$3.5 & 6.6$\stackrel{+3.6}{_{-3.0}}$  & 5.2$\stackrel{+5.1}{_{-3.0}}$  & 7.8$\pm$3.4  & 5.2$\pm$2.6      & 8.8$\pm$1.9 & 10$^{5}$\,K  \\
\ion{O}{v} &  $\gamma$    & 1.7$\pm$2.6 & 11.0$\pm$3.7  & 9.9$\pm$2.2  & 2.8$\stackrel{+3.0}{_{-3.2}}$  & 7.1$\pm$3.3  & 6.7$\pm$1.4    & km\,s$^{-1}$  \\
            &  $F$    & 30.5$\pm$2.8 & 33.8$\pm$3.0  & 26.6$\pm$2.6  & 39.1$\pm$3.3  & 31.7$\pm$2.9   & 31.8$\pm$1.3   & 10$^{-3}$\,erg\,cm$^{-2}$\,s$^{-1}$  \\
              &  $T$    & 3.5$\stackrel{+1.7}{_{-1.4}}$ & 13.9$\stackrel{+5.0}{_{-3.9}}$  & 2.0$\stackrel{+1.7}{_{-1.3}}$  & 10.6$\stackrel{+3.4}{_{-2.7}}$  & 7.8$\stackrel{+3.0}{_{-2.3}}$  & 7.2$\pm$1.2     & 10$^{5}$\,K  \\
\ion{N}{v} &  $\gamma$    & 3.9$\pm$1.2	 & 5.6$\pm$1.2  & 5.1$\pm$0.9  & 6.2$\pm$1.5  & 6.7$\pm$1.1      &  5.39$\pm$0.53     & km\,s$^{-1}$  \\
              &  $F^{1238.8}$    & 57.9$\pm$2.7 & 47.2$\pm$2.3  & 52.9$\pm$2.2  & 47.7$\pm$2.9  & 43.8$\pm$2.0	 & 48.8$\pm$1.2     & 10$^{-3}$\,erg\,cm$^{-2}$\,s$^{-1}$  \\
             &  $F^{1242.8}$    & 28.5$\pm$2.0 & 21.2$\pm$1.9 & 27.1$\pm$1.5	  & 26.1$\pm$2.2  & 19.2$\pm$1.5	   & 23.9$\pm$0.8   & 10$^{-3}$\,erg\,cm$^{-2}$\,s$^{-1}$  \\					   
              &  $T$    & 2.7$\stackrel{+1.5}{_{-1.4}}$ & 5.3$\stackrel{+1.5}{_{-1.8}}$ & 2.0$\stackrel{+1.2}{_{-0.9}}$ & 1.3$\stackrel{+2.4}{_{-1.0}}$ & 4.0$\stackrel{+1.1}{_{-1.5}}$  & 4.9$\pm$0.8  & 10$^{5}$\,K  \\
\ion{Fe}{xii} &  $\gamma$    & -3.2$\stackrel{+7.8}{_{-7.5}}$	 & -9.2$\stackrel{+9.8}{_{-8.4}}$  & 5.9$\stackrel{+4.1}{_{-4.7}}$  & -2.3$\stackrel{+12.3}{_{-11.0}}$  & 4.9$\stackrel{+8.1}{_{-7.0}}$	      &  3.5$\pm$3.2     & km\,s$^{-1}$  \\
             &  $F$    & 2.90$\stackrel{+0.70}{_{-0.66}}$	 & 2.51 $\pm$0.67   & 3.79$\pm$0.68  & 4.88$\stackrel{+1.30}{_{-1.13}}$ & 2.90$\stackrel{+0.74}{_{-0.69}}$ & 3.15$\pm$0.33    & 10$^{-3}$\,erg\,cm$^{-2}$\,s$^{-1}$  \\
              &  $T$    & 3.0$\stackrel{+1.8}{_{-1.5}}$  &   3.0$\stackrel{+3.8}{_{-2.2}}$   &  1.2$\stackrel{+1.7}{_{-0.9}}$   &  15.8$\stackrel{+24.0}{_{-9.2}}$   &  2.2$\stackrel{+3.2}{_{-1.5}}$  & 3.1$\stackrel{+1.2}{_{-1.0}}$      & 10$^{6}$\,K  \\	
\noalign{\smallskip} 
\hline
\noalign{\smallskip} 
\textbf{Stellar emission} &      &  &   &   &   &       &      &   \\ 
\noalign{\smallskip}
X-ray &  $F$   & - & 4.86$\stackrel{+0.47}{_{-0.35}}$ & 6.54$\stackrel{+0.14}{_{-0.11}}$ & 7.08$\stackrel{+0.14}{_{-0.13}}$   & 6.92$\stackrel{+0.13}{_{-0.12}}$  & 6.35$\stackrel{+0.13}{_{-0.10}}$   & erg\,cm$^{-2}$\,s$^{-1}$ \\    		
EUV      &  $F$   & - & 18.14$\stackrel{+4.45}{_{-4.32}}$ & 24.70$\stackrel{+2.98}{_{-2.81}}$ & 22.53$\stackrel{+2.87}{_{-2.30}}$   & 20.76$\stackrel{+3.45}{_{-2.58}}$  & 21.53$\stackrel{+1.75}{_{-1.55}}$   & erg\,cm$^{-2}$\,s$^{-1}$ 
\\
Photoionisation rate &  $\tau$(H\,{\sc i})   & - & 3.29$\stackrel{+0.67}{_{-0.58}}$ & 3.71$\stackrel{+0.59}{_{-0.51}}$ & 3.59$\stackrel{+0.47}{_{-0.45}}$   & 3.18$\stackrel{+0.53}{_{-0.44}}$  & 3.44$\stackrel{+0.28}{_{-0.25}}$   & 10$^{-7}$\,s$^{-1}$ 
\\
\\ 
\noalign{\smallskip} 
\hline
\noalign{\smallskip} 
\textbf{ISM}  &      &  &   &   &   &       &      &   \\ 
\noalign{\smallskip}
Component A &  $\gamma_{/\sun}$    & \multicolumn{6}{c}{-23.7$\stackrel{+1.2}{_{-1.7}}$}     & km\,s$^{-1}$ \\
          &  log$_{10}$\,$N$(H\,{\sc i})    & \multicolumn{6}{c}{18.34$\stackrel{+0.06}{_{-0.09}}$}      & cm$^{-2}$ \\
	          &  $T$    & \multicolumn{6}{c}{13074$\stackrel{+1137}{_{-1083}}$}      & \,K  \\	   
Component B &  $\gamma_{/\sun}$    & \multicolumn{6}{c}{-10.8$\stackrel{+2.2}{_{-2.4}}$}     & km\,s$^{-1}$ \\
            &  log$_{10}$\,$N$(H\,{\sc i})    & \multicolumn{6}{c}{17.93$\stackrel{+0.17}{_{-0.22}}$}      & cm$^{-2}$ \\
            &  $T$    & \multicolumn{6}{c}{5755$\stackrel{+2910}{_{-2802}}$}      & \,K  \\                                            
\hline
\hline
\end{tabular}
\begin{tablenotes}[para,flushleft]
Notes: $\gamma$ is the radial velocity of a model line centroid in the star rest frame, and $T$ its temperature. $F$ is the total flux at 1\,au from the star, in the model FUV lines, in the synthetic EUV spectra (62-912\,\AA), and in the model X-ray spectra (0.2-2.4\,keV = 5.2-62.0\,\AA). $\tau$(H\,{\sc i}) is the photoionisation rate of neutral hydrogen atoms corresponding to the mean XUV spectrum at 1\,au. $\gamma_{/\sun}$ is the radial velocity of a model ISM cloud relative to the Sun, log$_{10}$\,$N$(H\,{\sc i}) its column density of neutral hydrogen, and $T$ its temperature (assuming fixed turbulent broadening values, see text). 
  \end{tablenotes}
  \end{threeparttable}
\label{table:tab_paramsfit}
\end{table*}


\subsection{Analysis of the quiescent FUV stellar lines}
\label{sec:rec_FUVlines}

Stellar lines in the FUV provide useful information about the structure and emission of the chromosphere and transition region. We derived the properties of the \ion{Si}{iii}, \ion{O}{v}, \ion{N}{v}, and \ion{Fe}{xii} lines following the same approach as for the Lyman-$\alpha$ line (Sect.~\ref{sec:Ly_fit}). Models were fitted to the quiescent spectra, averaged over exposures identified as stable in each visit (Sect.~\ref{sec:search_var}). The average of the quiescent spectra over all visits further revealed the faint \ion{Si}{ii}\,$\lambda$1197.4 line. Voigt profiles were found to better model the bright \ion{Si}{iii} and \ion{N}{v} lines, while Gaussian profiles are sufficient for the \ion{Si}{ii}, \ion{O}{v}, and \ion{Fe}{xii} lines. We assumed that the model lines are thermally broadened, and allowed their centroid to vary in the star rest frame (defined by the GAIA radial velocity, assumed to trace the photosphere). We fitted a flat continuum level in the region of the \ion{Si}{ii}, \ion{Si}{iii} and \ion{N}{v} line, and a polynomial representing the Lyman-$\alpha$ red wing in the region of the \ion{O}{v} line. The \ion{Fe}{xii} line, blended with the \ion{N}{v}\,$\lambda$1239 line, was fitted together with the \ion{N}{v} doublet. We assumed a common temperature, damping parameter, and Doppler shift for the \ion{N}{v} lines. 

\begin{figure}    
\includegraphics[trim=1.8cm 2cm 3.4cm 8.4cm,clip=true,width=\columnwidth]{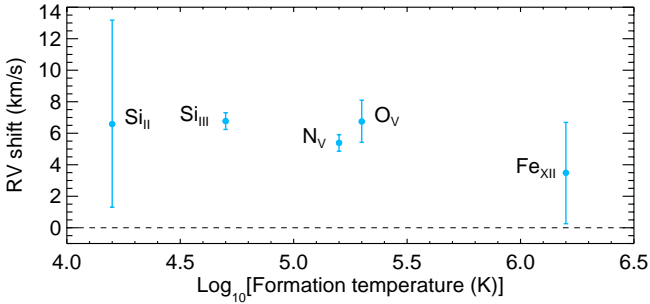}
\caption[]{Shifts of HD\,189733 emission lines relative to its photosphere radial velocity. Shifts have been derived from the quiescent lines averaged over all epochs. The Lyman-$\alpha$ line is not shifted with respect to the photosphere. Transition wavelengths associated to each line were taken from the NIST Atomic Spectra Database (\citealt{Kramida2016}). Line formation temperatures are from the Chianti v.7.0 database (\citealt{Dere1997}, \citealt{Landi2012})}
\label{fig:line_shifts}
\end{figure}

The best-fit properties for the quiescent lines in each visit, and for the lines averaged over all visits, are reported in Table~\ref{table:tab_paramsfit}. While the Lyman-$\alpha$ line is well aligned in the star rest frame (Sect.~\ref{sec:Ly_fit}), we detect significant redshifts for the \ion{Si}{iii}, \ion{N}{v}, and \ion{O}{v} lines (Fig.~\ref{fig:line_shifts}). This pattern is observed in the Sun (\citealt{Achour1995}, \citealt{Peter1999}) and other stars (e.g. \citealt{Linsky2012}), showing that the \textit{STIS} spectra of HD\,189733 are well calibrated and that the measured redshifts trace the structure of the transition region between the stellar chromosphere and corona. The amplitude and variation of HD\,189733 (P$_{\mathrm{rot}}\sim$8-10\,days) emission-line redshifts with formation temperature are consistent with those observed for stars rotating slower than $\sim$4\,days (\citealt{Linsky2012}). Namely, the Lyman-$\alpha$ line and chromospheric lines formed at low temperatures (e.g. \ion{Si}{ii}) do not show significant deviation from the photosphere velocity, while lines formed at logT$\approxsup$4.5 up to the transition region get increasingly resdhifted. The most redshifted lines of HD\,189733 are \ion{Si}{iii}, \ion{N}{v}, and \ion{O}{v}, formed at temperatures in between logT$\sim$4.7-5.3 that are expected to yield the maximum redshifts (\citealt{Linsky2012}). Redshifts are then expected to decrease with rising temperature, and coronal lines can even display blueshifts for logT$\approxsup$5.7. Although we do not have sufficient precision to draw a firm conclusion, the \ion{Fe}{xii} line is consistent with being less redshifted (Fig.~\ref{fig:line_shifts}). This pattern of deviations from the photosphere velocity can be explained by the heating of gas in the upper chromosphere, which propagates upward into the corona along the leg of a magnetic loop (emitting blueshifted lines), subsequently cools, and rains down along the other leg of the loop into the transition region (emitting redshifted lines). Interestingly the derived line temperatures (Table~\ref{table:tab_paramsfit}) are systematically larger than their expected formation temperatures (Fig.~\ref{fig:line_shifts}), which could possibly trace additional broadening due to the motion of the rising/falling gas. More information about this mechanism can be found in, e.g., \citet{Peter1999}, \citet{Hansteen2010}, \citet{Linsky2012} (see also \citealt{Bourrier2018_FUV} for possible signatures of planet-induced coronal rain in the star 55 Cnc).\\

\subsection{Analysis of the stellar X-ray spectrum}
\label{sec:X_spec}

\begin{figure}    
\includegraphics[trim=0cm 0cm 0cm 0cm,clip=true,width=\columnwidth]{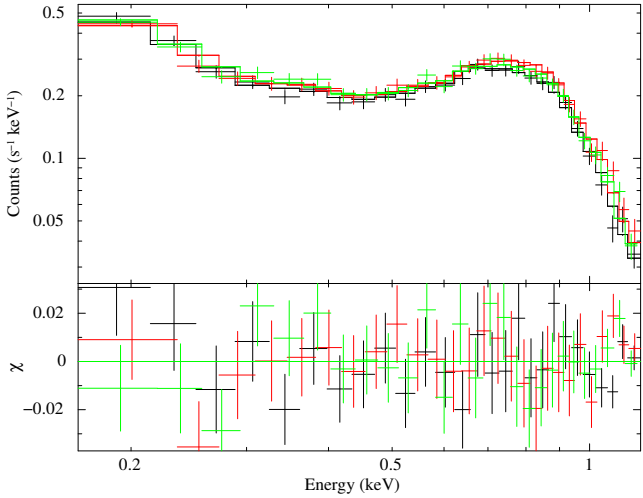}
\caption[]{X-ray spectra for the \textit{XMM-Newton} datasets (coloured points), together with their best-fit model for the spectra of HD\,189733 (coloured histograms), in the soft energy range uncontaminated by the background source. Visit C is in black, visit D is in red, and visit E is in green. The bottom panel show the residual between the measured spectra and their fit.}
\label{fig:Xspec}
\end{figure}

The \textit{XMM-Newton EPIC-pn} spectra extracted in Sect.~\ref{sec:xray_obs} for Visit C, D, and E were analysed using \textsc{xspec} 12.9.1p \citep{Arnaud1996}. The X-ray spectra for these three visits, displayed in Fig.~\ref{fig:Xspec}, are very similar in shape and flux. They are reasonably soft, typical of moderately active late-type stars a few Gyr old, including strong line emission between 0.6 and 0.9\,keV due primarily to oxygen and the  L-shell transitions of iron.

We used \textsc{apec} models (\citealt{smith2001}) to fit the X-ray spectra, limiting the fit to energies below 1.2\,keV to avoid contamination by the background source (Sect~\ref{sec:xray_obs}). We found that at least three temperature components were required to reproduce the spectra (k\,T$_{1}$ = 0.147$^{+0.059}_{-0.007}$\,keV, k\,T$_{2}$ = 0.345$^{+0.031}_{-0.032}$\,keV, k\,T$_{3}$ = 0.724$^{+0.034}_{-0.090}$\,keV). Although we stress that this should be thought of as an approximation to a plasma with a continuous range of temperatures. Each temperature component was linked across the three observations, thereby forcing the spectral shape to remain the same. However, we did allow the total emission measures, and therefore fluxes, to change. An interstellar absorption term was included for completeness by using the \textsc{tbabs} model (\citealt{wilms2000}), though its contribution to the results was negligible given the relatively close proximity of HD\,189733 to Earth (Table~\ref{tab:param_sys}).

Using fixed Solar abundances across all species (\citealt{Caffau2011}), we were not able to attain a statistically acceptable fit. Freeing up elements (C, N, O, Ne, Fe) relevant for the first ionisation potential (FIP) effect (e.g., \citealt{Feldman1992}, \citealt{Laming2015}) yielded a far superior fit. We obtained a coronal Ne/Fe value of 7.9, a value in between quiet and very active K0/K1 stars \citep{Wood2012,Laming2015}. Additionally, we estimate from the coronal and photospheric abundances $F_{\rm bias}$ = log$_{\rm 10}$(X / Fe)$_{\rm cor}$ - log$_{\rm 10}$(X / Fe)$_{\rm phot}$, where X is abundance of the high FIP species being tested. Our final value of $F_{\rm bias}$ is $0.65^{+0.15}_{-0.07}$, obtained by averaging across the four species we freed up, and is indicative of a relatively strong inverse FIP effect in HD\,189733. This result disagrees with that of \citet{Poppenhaeger2013}, who found an $F_{\rm bias}$ of -0.41 in their fit to six \textit{Chandra} observations when freeing up only Ne, O, and Fe. However, our abundances for N, which exhibits the greatest inverse FIP effect, and Fe are independently corroborated in the differential emission measure (DEM) fit to the high-excitation FUV lines in Sec.~\ref{sec:EUV_spec}. Our resultant X-ray fluxes in the 0.2-2.4\,keV band are given in Table~\ref{table:tab_paramsfit} and shown in Fig.~\ref{fig:Fig_Time_evol_fluxes}. 

We forced the fit to the \textit{Swift} data in Visit B to have the same temperatures and abundances as the \textit{XMM-Newton} fit, given the limited number of counts. The normalisations, and hence fluxes, were allowed to change. We fitted independently the spectra for the flaring and non-flaring times of HD\,189733. In the 0.2-2.4\,keV band we derive a flux of $\left( 4.86^{+0.47}_{-0.35} \right)$\,erg\,s$^{-1}$\,cm$^{-2}$ for the non-flaring spectrum at 1\,au from the star, lower than the three \textit{XMM-Newton} epochs. In-flare, the flux rises to $\left( 12.0^{+3.9}_{-0.9} \right)$\,erg\,s$^{-1}$\,cm$^{-2}$, unsurprisingly considerably higher than the quiescent flux in all three \textit{XMM-Newton} observations.

\subsection{Reconstruction of the stellar EUV spectrum}
\label{sec:EUV_spec}

Most of the stellar EUV spectrum is not observable from Earth because of ISM absorption. We therefore reconstructed the entire XUV spectrum up to 1600\,\AA\, using the DEM retrieval technique described in \citet{Louden2017}. This reconstruction is based on the quiescent X-ray flux for the coronal region (from the data shown in Sect.~\ref{sec:search_var}), and on the flux derived for the intrinsic FUV stellar lines (Sect.~\ref{sec:rec_FUVlines}) for the chromosphere and transition region. The XUV spectra could thus be reconstructed for Visits B to E. 

\begin{figure}    
\includegraphics[trim=0cm 0cm 0cm 0cm,clip=true,width=\columnwidth]{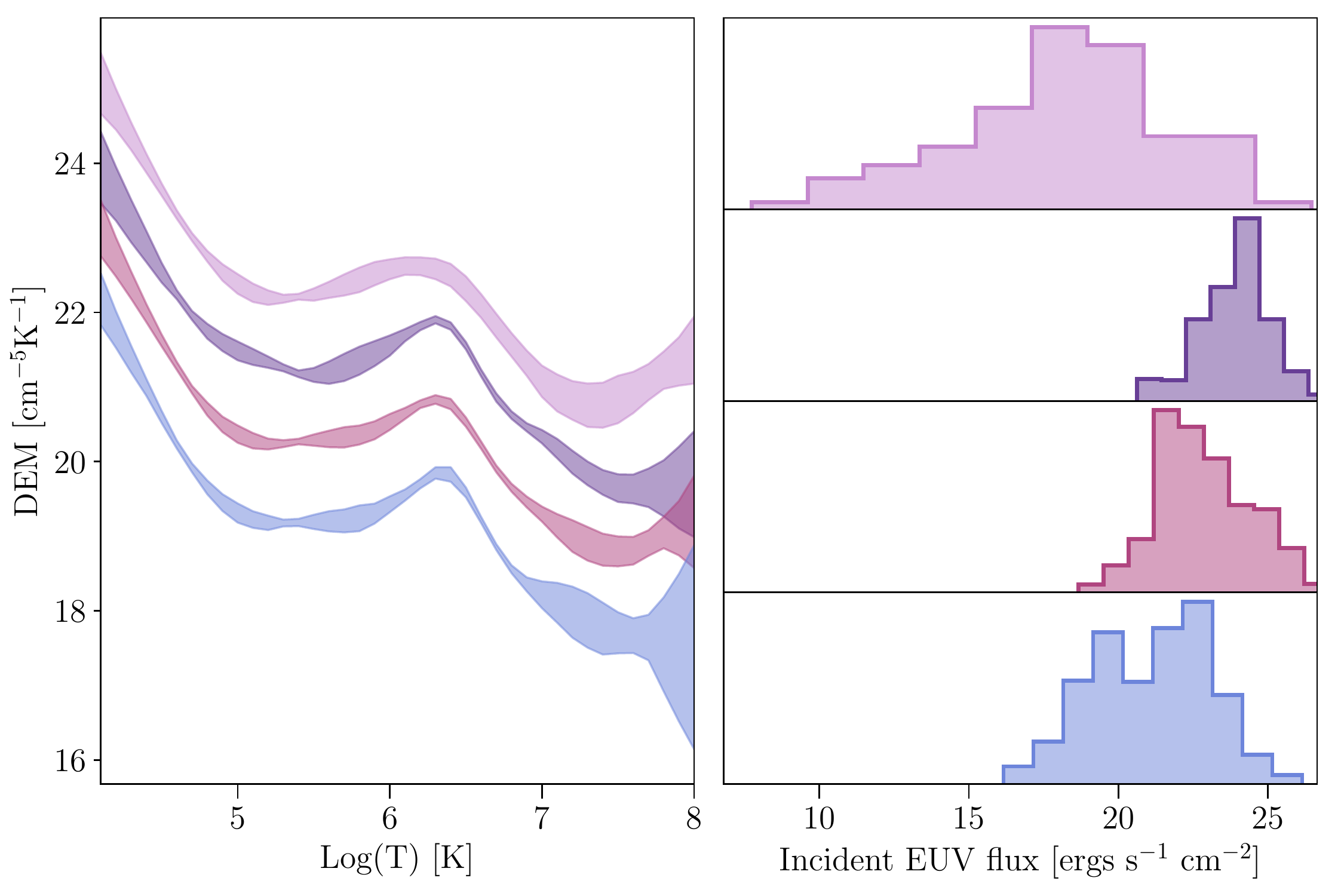}
\caption[]{\textit{Left panel:} The 1\,$\sigma$ credible region for the DEM's for Visits B to E (from top to bottom) with offsets for clarity. The shape of the DEM remains consistent between the 4 visits, except for the very high temperatures where the constraints from the data are weaker. \textit{Right panel:} The posteriors of the EUV (62-920 \AA) flux at 1\,au for the 4 nights, in the same order. The constraints are weakest in Visit B due to the lower SNR of the \textit{Swift} data, demonstrating the necessity of wide coverage for accurate reconstructions.}
\label{fig:dems}
\end{figure}

Due to the high SNR of the X-ray spectra we chose not to use Chebyshev Polynomials to calculate the shape of the DEM's, instead using a regularised inversion approach, as in \citet{Hannah2012}, otherwise the technique is the same as described in \citet{Louden2017}. We found that a regularisation parameter of 100 gave the best balance between model complexity and fit to the data. Initially, the abundances were set to solar photospheric values (\citealt{Caffau2011}) for the UV lines and solar coronal (\citealt{Schmelz2012}) for X-ray flux. This resulted in a poor fit to the \ion{O}{v} and \ion{N}{v} lines, which are formed at a similar characteristic temperature of $\sim$10$^{5.3}$\,K, and the \ion{Fe}{xii} line, formed at $\sim$10$^{6.3}$\,K. We then fit again, allowing the nitrogen and iron abundances to vary. We found consistently in the four visits that the best fitting values for the nitrogen and iron abundances were respectively 3.8 and 0.6 times the values derived by \citealt{Caffau2011}. This is not an unexpected result, due both to the gross differences in abundances between stars, and also the potential effects of the FIP and inverse FIP effects in modifying these abundances in the star's upper atmosphere. We then repeated the fit for each of the four visits with these values fixed to generate our final DEM and spectra. With the abundances thus modified the fits to the lines were significantly improved. The flux of each line was recovered on each night to within 1.5$\sigma$, and the X-ray flux was recovered with consistent values to those found in Sect.~\ref{sec:X_spec}. The final DEM's, as well as the posteriors for the total EUV fluxes at 1\,au, are plotted in Fig.~\ref{fig:dems}, with the corresponding best-fit values given in Table~\ref{table:tab_paramsfit}. The generated XUV spectra for the four visits are available online as machine readable tables.

The synthetic XUV spectra yield an average photoionisation lifetime of about 33.6 days for neutral hydrogen at 1\,au from HD\,189733, which corresponds to 50\,min at the orbital distance of HD\,189733b (details on the calculation can be found in \citealt{Bourrier2017_HD976}). Implications for the structure of the planetary exosphere are discussed in Sect.~\ref{sec:interp}. The lifetime at 1\,au ranges within its uncertainties between 29 and 43 days in Visit C, which is close to the range $\sim$6.5-25 days derived by \citealt{Bourrier_lecav2013} from the fit to the Lyman-$\alpha$ transit in this visit. As the atmospheric mass loss of neutral hydrogen correlates positively with  its photoionisation rate in this fit, the larger lifetimes we obtained suggest that the neutral hydrogen loss from HD\,189733b in Visit B is at the lower end of the range derived by \citet{Bourrier_lecav2013} ($\sim$10$^{9}$\,g\,s$^{-1}$ or lower). \\

\subsection{Temporal evolution of the stellar XUV spectrum}
\label{sec:evol_XUV}

\begin{figure}    
\includegraphics[trim=0cm 0cm 0cm 0cm,clip=true,width=\columnwidth]{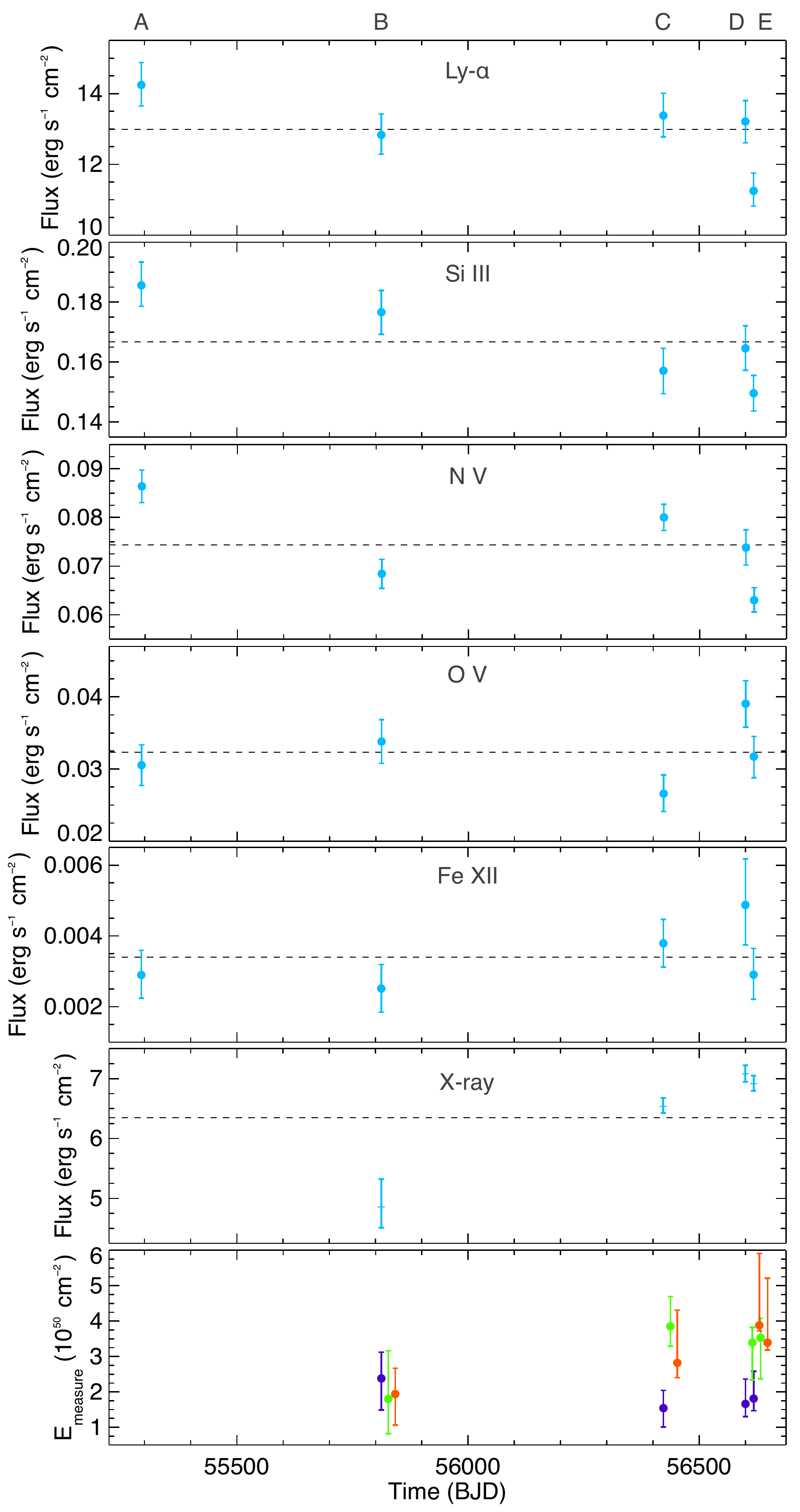}
\caption[]{Emission from the chromosphere and corona of HD\,189733 as a function of time. The upper panels show the total flux at 1\,au from the star, in the theoretical intrinsic FUV lines fitted to the \textit{HST} spectra (lines are ordered by increasing formation temperature from top to bottom) and in the 0.2-2.4\,keV X-ray band (derived from the \textit{Swift} and \textit{XMM-Newton} spectra). The dashed black line is the mean flux over all epochs. The bottom panel shows the emission measures associated with the low (purple), medium (green), and high (orange) temperatures in the fitted X-ray spectra (slightly offset in each epoch for clarity).}
\label{fig:Fig_Time_evol_fluxes}
\end{figure}

Overall the quiescent emission of HD\,189733 in the observed FUV lines remains quite stable over the two years and a half covered by our visits (Fig.~\ref{fig:Fig_Time_evol_fluxes}). We note, however, two interesting features. First, low-temperature lines (Lyman-$\alpha$, \ion{Si}{iii}, \ion{N}{v}) became weaker overall between Visit A and Visits D+E, while the flux in high-temperature lines (\ion{O}{v}, \ion{Fe}{xii}) and X-rays increased. This could possibly trace a decrease in the chromospheric activity of HD\,189733 from 2010 to 2013, while its corona became more active. The X-ray flux, in particular, increased significantly from visits B to the next visits. Over the same period the emission measures associated with each of the three X-ray temperature components remained consistent within their uncertainties, but the emission measures for the medium and high temperatures show a marginal increase. The second notable feature is the decrease in flux from Visit D to Visit E visible in all spectral bands, albeit stronger at low energies.\\

Both the increase in X-ray flux up to Visit D, and the global decrease in the emission of HD\,189733 between visits D and E, correspond well to the evolution of the magnetic field over 9 years reported in \citet{Fares2017}. In this study, we showed that the mean intensity of the magnetic field increased from 18\,G to 42\,G between 2006 and September 2013 (just before Visit D), before dropping sharply to 32\,G in September 2014 (after Visit E) and becoming more toroidal.\\

The intrinsic Lyman-$\alpha$ line shows little variations in amplitude from Visit A to D, and keeps nearly the same shape (Fig.~\ref{fig:La_stellar_lines}). Once rescaled to the same amplitude, the line profiles show nearly no variations in the wings and the peaks, and only slight variations in the depth of the self-reversal. In Visit E however, the line not only becomes weaker but also shows a shallower self-reversed core,  which likely traces changes in the structure of the transition region associated with the aforementioned variations in the structure of the magnetic field and corona. \\

\section{Interpretation of the FUV absorption signatures}
\label{sec:interp}

Our goal in this section is to qualitatively explore scenarios that could explain the observed FUV variations. Detailed modelling of HD\,189733b upper atmosphere will be carried out in following papers of the MOVES series. \\

\subsection{\ion{Si}{iii} and \ion{N}{v} lines}

Flux decreases were observed in the \ion{Si}{iii} line before and/or during the transit in Visits A and B. In Visits B and E the weaker line of the \ion{N}{v} doublet showed variability, even though the brighter doublet line remained stable.

We used the synthetic stellar spectra derived in Sect.~\ref{sec:EUV_spec} to calculate photoionisation lifetimes at the orbital distance of HD\,189733b. On average it takes $\sim$35\,h for a neutral nitrogen atom to be ionised into N$^{4+}$ (lifetimes associated with successive ionisations are 7.4+19.9+86.2+1989.2\,min). It would thus be unlikely for nitrogen atoms escaping the planet to remain in its vicinity long enough to be photoionised four times, which is in agreement with the non-detection of absorption in the brightest line of the \ion{N}{v} doublet. The variations observed both in emission and absorption in the weaker line of the doublet (Sect.~\ref{sec:VB}, ~\ref{sec:VE}), if they are not artefacts due to the lower SNR, likely trace stellar activity leading to density variations in the dominant emission regions of the corona.

In contrast, it takes $\sim$2\,h for a neutral silicon atom to be ionised into Si$^{2+}$ (successive lifetimes are 0.5+117.5\,min). Ionised silicon atoms escaped from the planet could thus remain in its vicinity and absorb the flux in the stellar \ion{Si}{iii} line near the transit, before they are carried away by stellar wind, radiation pressure, or magnetic interactions. However the escape of planetary silicon atoms would not explain why they were only detected in Visits A and B, and why they yield absorption only before/during the planetary transit in those visits. The change in the stellar spectral energy distribution (SED) from Visit B to E, which results in respective photoionisation lifetimes of 2.3, 1.7, 1.9, and 2.0\,h, does not explain these differences. We also note that the non-detection of Lyman-$\alpha$ absorption in the same time windows as the \ion{Si}{iii} signatures in Visits A and B shows that the absorber is in a high ionisation state. Another scenario proposed by \citet{Bourrier2013} is that the encounter of the stellar wind with the planet magnetosphere leads to the formation of a shock in which metals like silicon could quickly get ionised in a dense front ahead of the planet. The orientation and stand-off distance of this bow-shock would vary over time depending on the velocity of the stellar wind and the strength of the planet magnetic field (e.g. \citealt{Llama2011}, \citealt{Vidotto2010}). In this scenario, absorption of the \ion{Si}{iii} line just before and during the transit in Visit A would imply that the shock is facing the star more than in Visit B, when it is only visible before the transit. This could result from a faster stellar wind in Visit A (see the extreme cases of dayside- and ahead-shocks in \citealt{Vidotto2010}). The non-detection of the shock in subsequent visits could be linked to inhomogeneities in the density of the stellar wind along the planetary orbit (\citealt{Llama2013}, \citealt{Kavanagh2019}), resulting in a shock not dense enough to absorb the stellar lines. Alternatively it could be linked with the significant increase in X-ray emission, which traces an evolution of the coronal and stellar wind properties. For example higher coronal and wind temperatures could result in a more energetic shock that would ionise silicon atoms to higher levels than Si$^{2+}$. We note that silicon atoms escaping from HD\,189733b are not required in this scenario, as the ionised population within the shock could originate from the stellar wind.\\


\subsection{Lyman-$\alpha$ line}

In the following subsections we show how the absorption signatures observed in the Lyman-$\alpha$ line in Visits B, D, and E could arise from the upper atmosphere of HD\,189733b, and how their variable properties and the non-detection in Visit A could be linked to the evolution of the stellar SED and stellar wind. 

\subsubsection{Visits B and D} 

\begin{figure}     
\includegraphics[trim=3.5cm 0cm 0cm 0cm,clip=true,width=\columnwidth]{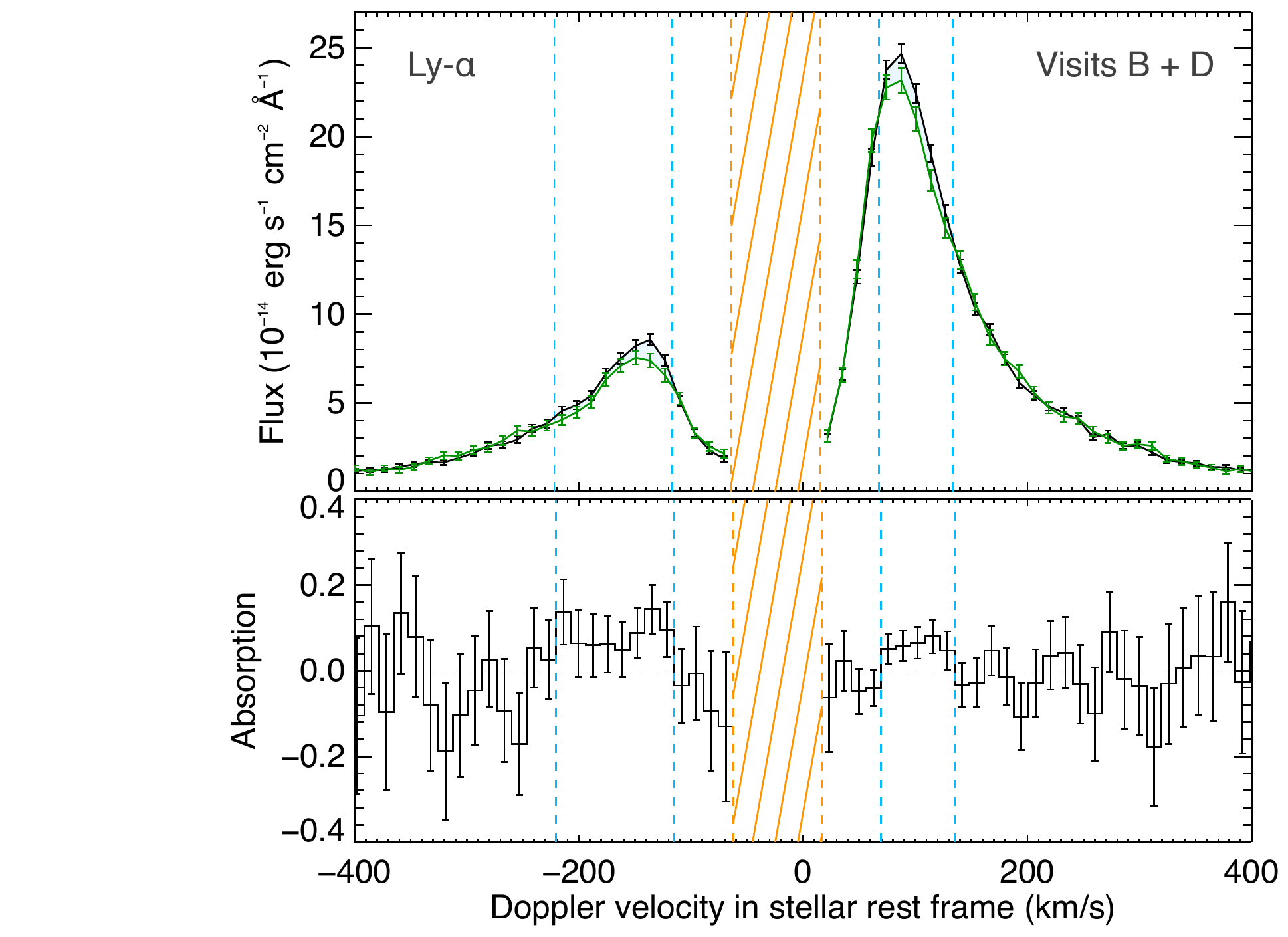}
\caption[]{\textbf{Top panel:} Stellar Lyman-$\alpha$ line averaged over visits B and D. Spectral ranges showing absorption during the transit (green spectrum) compared to pre-transit orbits (black spectrum) are highlighted in blue. The dashed orange region is contamintaed by the airglow in Visit D. \textbf{Bottom panel:} Absorption spectrum defined as the relative flux ratio between the in- and pre-transit spectra in Visit D.}
\label{fig:flux_sp_V2V4}
\end{figure}

As mentioned in Sect.~\ref{sec:VD} these two visits show the most similarities. To highlight this point we average their pre- and in-transit Lyman-$\alpha$ line spectra in Fig.~\ref{fig:flux_sp_V2V4}. The signature at the peak of the red wing is more clearly revealed, although the absorption depth in excess of the planetary continuum remains marginal (3.6$\pm$1.7\% between 68 and 134\,km\,s$^{-1}$). The absorption signature in the blue wing, which extends over larger velocities in Visit B, yields an average excess absorption of 6.4$\pm$2.3\% between -222 to -117\,km\,s$^{-1}$.

If confirmed, absorption in the red wing could possibly arise from the extended thermosphere of HD\,189733b (\citealt{BJ2008}, \citealt{Guo2016}). The similar levels of EUV irradiation and hydrogen photoionisation rates in Visits B and D (Table~\ref{table:tab_paramsfit}) would lead to similar structures for the layer of neutral hydrogen, explaining the repeatability of the absorption signature. In contrast to \citet{Guo2016}, however, we do not believe that the red wing absorption can arise entirely from the Lorentzian wings of the thermospheric absorption profile (the so-called damping wings). In this case, absorption should increase toward low velocities in the planet rest frame (see e.g. \citealt{Bourrier2018_GJ3470b}), whereas no absorption is detected in Visit B or D between $\sim$30 and 70\,km\,s$^{-1}$ (Fig.~\ref{fig:La_stellar_lines}). 

Because of ISM absorption, blueshifted velocities lower than about -100\,km\,s$^{-1}$ cannot be probed, preventing us from searching for the symmetrical signature expected from an extended thermosphere. The absorption signatures measured in the blue wing, at much larger velocities than in the red wing, cannot be due to the extended thermosphere. They also cannot be explained by radiation pressure, which requires more than 7\,hours to accelerate neutral hydrogen atoms escaping the planet to a terminal radial velocity of $\sim$140\,km\,s$^{-1}$ (\citealt{Bourrier_lecav2013}). The observed signatures could be explained by charge-exchange between the stellar wind and the planetary exosphere, which would create a population of energetic neutral hydrogen atoms (\textit{ie}, former stellar wind protons that got neutralised) moving with the velocity distribution of the stellar wind (\citealt{Lecav2012}, \citealt{Bourrier_lecav2013}). In this scenario, the different velocity ranges measured in Visits B (-220.3 to -128.1\,km\,s$^{-1}$) and Visit D (-156.1 and -116.5\,km\,s$^{-1}$) could trace variations in the stellar wind properties. As shown in \citet{Bourrier_lecav2013}, the neutralised stellar wind protons are expected to trail the planet over a short distance before they are photoionised (lifetime$\sim$50\,min), which would further explain the absence of absorption after the transit in Visit B and at egress in Visit D (Sect.~\ref{sec:VB}, ~\ref{sec:VD}). We thus favour interactions between the wind of HD\,189733 and the exosphere of HD\,189733b as the source for the measured Lyman-$\alpha$ blueshifted absorption signatures.\\


\subsubsection{Visit A} 

\citet{Lecav2012}, \citet{Bourrier_lecav2013} proposed that the atmospheric escape detected in Visit B was linked to the increased energy input and/or different stellar wind properties associated with the flare that occurred about 8 hours before the transit. However, \citet{Chadney2017} showed that the short duration and energy spectrum typical of a flare on HD\,189733, while increasing the total atmospheric loss from the planet, enhances neutral hydrogen loss by at most a factor of 2. This is not sufficient to explain the increase in absorption depth from Visit A to Visit B (\citealt{Bourrier_lecav2013}). Furthermore if the observed blueshifted absorption signatures arise from stellar wind protons associated with a flare, the neutralised proton tail would have had to form and remain neutral within a limited time window after the flare, suggesting the stellar wind would only be enhanced for a short duration. It seems unlikely that we would have observed the transit of a tail in both Visits B, D, and E (see below) at the right time a few hours after a stellar flare.

Another possibility to explain the difference between Visit A and subsequent visits is the change in the stellar SED. If the X-ray emission followed the same trend as in other visits, it was significantly lower in Visit A (Fig.~\ref{fig:Fig_Time_evol_fluxes}), likely decreasing the global extension of the thermosphere and its mass loss rate. Meanwhile Visit A shows the largest flux of all visits in the Lyman-$\alpha$ and \ion{Si}{iii} lines, implying that the EUV irradiation on the planet was highest in this epoch and that the hydrogen layer within the thermosphere was more ionised. \citet{Guo2016} suggested that an increase in the ratio F(50-400\,\AA)/F(50-900\,\AA) by a factor 2 from Visit A to Visit B would have made the thermosphere dense enough in neutral hydrogen for its damping wings to become detectable in the red wing of the Lyman-$\alpha$ line in Visit B. However we note that a change in the thermospheric structure does not explain by itself the variations observed in the blue wing of the line, where absorption signatures are measured at velocities too high to arise from the thermosphere. It is the reduced escape rate of neutral hydrogen from the thermosphere, combined with the higher photoionisation of escaping hydrogen atoms, which could have limited the abundance of neutral hydrogen atoms arising from charge exchange with the exosphere in Visit A. For example \citet{Bourrier_lecav2013} showed that photoionisation rates at least 3 times larger in Visit A than in Visit B, with similar escape rates of neutral hydrogen, would have prevented the formation of a detectable neutralised proton tail detectable in the blue wing of the Lyman-$\alpha$ line.\\


\subsubsection{Visit E} 

There are strong similarities between Visit E and Visits B+D. In the two cases, absorption signatures are measured during the transit, in both wings of the Lyman-$\alpha$ line and from about the same velocities (70\,km\,s$^{-1}$ in the red wing, -110\,km\,s$^{-1}$ in the blue wing), while the core of the line remains stable. On the other hand, signatures are deeper in both wings of the line in Visit E, and extend up to larger velocities than in Visits B+D.\\

Interestingly, Visit E shows the lowest flux of all visits in low-energy chromospheric lines and one of the largest X-ray emission, in particular at soft energies mostly responsible for heating in the upper atmosphere (Fig.~\ref{fig:Fig_Time_evol_fluxes}, Table~\ref{table:tab_paramsfit}). This evolution in the stellar SED could have led to a dramatic change in the upper atmosphere of HD\,189733b. Based on the calculations by \citet{Owen2012}, HD189733b is close to the limit where the escape regime transitions from being EUV- to X-ray driven (see their Fig.~11, with a = 0.032\,au, $\rho$ = 0.8\,g/cm$^{3}$, L$_\mathrm{x}(\lambda<100$\AA$)\sim$3$\times$10$^{28}$\,erg\,s$^{-1}$ in Visit E). This transition might have occurred in Visit E, leading to larger mass loss in the X-ray driven regime (\citealt{Owen2012}). The low photoionisation rates in this epoch would have further increased the abundance of neutral hydrogen in the escaping outflow, amplifying absorption signatures in this epoch. The similarities of the absorption signatures in Visit B, D, and E at low velocities suggest that the dynamics of the neutral hydrogen layer close to the planet remains controlled by the same mechanism. The larger velocities of the red wing signature in Visit E suggest that we probe neutral hydrogen gas moving farther from the planet, possibly a stream accreting toward the star revealed by the enhancement in neutral hydrogen abundance (\citealt{Lai2010}; \citealt{Lanza2014}, \citealt{Matsakos2015}; \citealt{Strugarek2016}). Such a stream could yield transit signatures in the Balmer lines at even larger distance from the planet, as suggested by the detection of pre- and post-transit absorption in ground-based optical observations (see \citealt{Cauley2017} and references within). The larger velocities of the blue wing signature in Visit E might trace larger local velocities of the stellar wind at the orbit of the planet. Residual absorption might have been detected in additional post-transit observations from a tail trailing the planet, in contrast to Visit B. \\

\section{Discussions}
\label{sec:conc}

The corona of HD\,189733 became more active over the course of our observations while chromospheric activity decreased, which likely led to substantial changes in the planetary environment. In particular the variations observed in X-ray emission correlate well with the evolution of the stellar magnetosphere described in \citet{Fares2017}. The stellar wind properties in Visits A and B, when the stellar magnetic and coronal activity was reduced, could have favoured the formation of a dense population of Si$^{2+}$ atoms in a bow-shock ahead of the planet, responsible for the pre- and in-transit absorption measured in the \ion{Si}{iii} line in those visits. Meanwhile, we surmise that a lower X-ray irradiation and larger photoionisation of the planet in Visit A could have limited the extension and neutral content of its upper atmosphere, explaining why no Lyman-$\alpha$ transit was detected in this epoch. In subsequent visits, the change in stellar SED may have increased the abundance of neutral hydrogen in the thermosphere, which could be partly responsible for the absorption signatures detected in Visits B, D, and E at low velocities in the red wing of the Lyman-$\alpha$ line. The corresponding increase in neutral hydrogen escape would further explain the absorption signatures observed in those visits at high velocities in the blue wing of the line, arising from charge-exchange between the variable stellar wind and the (neutral) hydrogen exosphere. A sharp change in the structure of the star magnetosphere and its high-energy emission from Visit D to Visit E might then have led to a dramatic change in the evaporation regime of the planet, sustaining a much larger neutral hydrogen loss. In any invidual epoch, no transit signatures were detected in the \ion{N}{v} and \ion{O}{v} lines, or in the X-rays. 

Based on these results, we make the following predictions:
\vspace{-8pt}
\begin{enumerate}
\item epochs of low EUV emission and high X-ray emission enhance the abundance of neutral hydrogen in the thermosphere and the exosphere of HD\,189733b, and are thus the most favourable to probe the upper atmosphere via Lyman-$\alpha$ transit spectroscopy.
\item absorption from the upper atmosphere is maximal during the time window of the planetary transit. 
\item epochs of low X-ray emission, possibly associated with a less energetic stellar wind, lead to the formation of a bow-shock enriched in low-ionisation species and with a different orientation.
\end{enumerate} 
Future observations of the planetary environment could thus be planned based on the predicted stellar activity level, and should monitor the stellar X-ray emission while searching for the transit of the planet upper atmosphere and escaping outflow in FUV lines and other potential tracers.\\

Overall, the XUV irradiation of HD\,189733b is high enough in all epochs for the upper atmosphere to be extended and escaping (e.g. \citealt{Owen2012}). The detection of repeatable absorption signatures in three independent epochs from an extended but compact helium layer (\citealt{Salz2018}) further suggests that the lower regions of the extended thermosphere are stable over time. However, our study shows how variations in the stellar XUV spectrum and wind properties could influence the density of neutral hydrogen in the upper thermosphere and in the exosphere, resulting in the variability of their observed Lyman-$\alpha$ transit signatures. It is worth noting that the Lyman-$\alpha$ blueshifted absorption signatures would trace the local conditions of the stellar wind. Observations of the solar wind (e.g. \citealt{McComas2008}) and numerical simulations of stellar winds (\citealt{Llama2013}, \citealt{Kavanagh2019}) have shown that winds are not homogeneous, with streams of high and low velocity material coexisting at any given epoch. These inhomogeneities are linked to the topology of stellar magnetic fields, which for HD\,189733b is known to be complex and evolve over time (\citealt{Fares2017}). This variability of HD\,189733b upper atmosphere was hinted by previous unresolved observations of the Lyman-$\alpha$ line with \textit{HST}/\textit{ACS} (\citealt{Lecav2010}), which showed consistent excess atmospheric absorption depths of 5.2$\pm$ 1.5\% on 2007-06-10 and 3.2$\pm$1.7\% on 2007-06-18/19 (four transits later), but no excess in a third visit in April 2008. We note, however, that a flare observed during the April 2008 visit might have contaminated the observations. Our study revealed another flare from the primary star that occurred during the transit in Visit C, with a FUV-signature only. Flares ocuring repeatedly after the secondary eclipse have been proposed as a signature of SPI between HD\,189733b and its star (\citealt{Pillitteri2010,Pillitteri2011,Pillitteri2014}). In contrast, despite many transit observations of HD\,189733b this is only the third time that a flare is observed during the primary transit (the first was measured in the FUV by \citealt{Lecav2010}, the second in optical chromospheric lines by \citealt{Klocova2017}), making it unlikely that they are induced by the planet. \\

\section{Conclusions}
\label{sec:conc}

This paper is part of the MOVES collaboration, which aims at characterising the environment of the hot Jupiter HD\,189733b and its star via multi-wavelength observations obtained contemporaneously in different epochs. In MOVES I (\citealt{Fares2017}) we used optical spectropolarimetry to reconstruct the 3D stellar magnetosphere and study its evolution over several years. These results were used in MOVES II (\citealt{Kavanagh2019}) to model the wind of the host star and study its influence on the planetary radio emission. In the present paper (MOVES III), we combined FUV and X-ray observations to characterise the stellar high-energy emission, to search for transit signatures from the planet upper atmosphere, and to study their evolution in five epochs from 2010 to 2013.\\

Transit signatures are measured in the stellar Lyman-$\alpha$ line in three epochs, and in the \ion{Si}{iii} line in two epochs. These signatures could be related to the evolution of the stellar high-energy emission and stellar wind (MOVES II), linked to the evolution of the stellar magnetosphere (MOVES I). Our analysis thus confirms the evaporation of HD\,189733b and its temporal variability, and supports the presence of a bow-shock ahead of the planet.

Knowledge of the stellar irradiation is key to our understanding of close-in planet atmospheres. Their photochemistry and stability is influenced by XUV photons, making critical the use of realistic stellar spectra in atmospheric models. Previous studies of HD\,189733b have used a variety of approaches, including the solar EUV spectrum, stellar proxys like $\epsilon$ Eridiani, or theoretical spectra (see discussion in \citealt{Guo2016}). Here we combined measurements of the soft X-ray emission and chromospheric/transition region lines from HD\,189733 to constrain a model of the stellar atmosphere and reconstruct its entire XUV spectrum in four epochs. These synthetic spectra, which extend up to 1600\,\AA\,, are available online. The Lyman-$\alpha$ line of HD\,189733 alone represents half of the flux emitted in the entire EUV domain. It is particularly important for atmospheric chemistry (e.g. \citealt{Miguel2015}) but also for the structure of the planetary exosphere affected by radiation pressure. We thus include our reconstructed profiles for the intrinsic Lyman-$\alpha$ line of HD\,189733 in its synthetic spectra. This reconstruction further allowed us to characterise the ISM properties toward HD\,189733, revealing at least two clouds along the LOS.

Future studies of the MOVES collaboration will combine the stellar wind models from MOVES II with the synthetic stellar XUV spectra derived here to inform detailed models of HD\,189733b upper atmosphere, and determine the mechanisms and physical properties responsible for the observed FUV signatures.\\

\section*{Acknowledgements}

We thank the referee for their attentive review of our paper. We thank D.K.~Sing for his help and insights into the investigation of the \textit{HST} jitter. V.B. acknowledges support by the Swiss National Science Foundation (SNSF) in the frame of the National Centre for Competence in Research ``PlanetS'', and has received funding from the European Research Council (ERC) under the European Union's Horizon 2020 research and innovation programme (project Four Aces; grant agreement No 724427). P.J.W, G.K and T.L acknowledge support from the UK Science and Technology Facilities Council (STFC) under the consolidated grants ST/L000733/1 and ST/P000495/1. A.L.E. acknowledges support from CNES and the French Agence Nationale de la Recherche (ANR), under programme ANR-12-BS05-0012 Exo-Atmos. A.A.V. acknowledges funding from the Irish Research Council. This work is based on observations made with the NASA/ESA Hubble Space Telescope, obtained at the Space Telescope Science Institute, which is operated by the Association of Universities for Research in Astronomy, Inc., under NASA contract NAS 5-26555. This work is based on observations obtained with \textit{XMM-Newton}, an ESA science mission with instruments and contributions directly funded by ESA Member States and NASA. This research has made use of data obtained from NASA's \textit{Swift} satellite. This work has made use of data from the European Space Agency (ESA)
mission {\it Gaia} (\url{https://www.cosmos.esa.int/gaia}), processed by
the {\it Gaia} Data Processing and Analysis Consortium (DPAC,
\url{https://www.cosmos.esa.int/web/gaia/dpac/consortium}). Funding
for the DPAC has been provided by national institutions, in particular
the institutions participating in the {\it Gaia} Multilateral Agreement.

\let\mnrasl=\mnras
\bibliography{biblio} 
\bsp
\label{lastpage}

\end{document}